\documentclass[letterpaper,nofootinbib,aps,prd,onecolumn,superscriptaddress,floats,amsfonts,amssymb,12pt]{revtex4}

\usepackage{amsmath,yhmath}
\usepackage{comment}
\usepackage{graphicx}
\usepackage{epstopdf}
\usepackage[active]{srcltx}
\usepackage{bm}
\usepackage[]{latexsym}
\usepackage[margin=0.5cm]{caption}

\usepackage{color}

\usepackage{float}

\usepackage{hyperref}

\usepackage{lineno}
\usepackage{enumitem}
\begin{document}
%\linenumbers

\restylefloat{figure}

%Revised
\newcommand*{\st}[1]{\textbf{\color{red}{*} #1 * }}

\title{Testing MSW effect in supernova explosion with neutrino event rates}

\author{Kwang-Chang Lai}
%\email{kcl@mail.cgu.edu.tw}
\affiliation{Center for General Education, Chang Gung University, Kwei-Shan, Taoyuan, 333, Taiwan}
%\affiliation{Leung Center for Cosmology and Particle Astrophysics (LeCosPA), National Taiwan University, Taipei, 106, Taiwan}

\author{C. S. Jason Leung}
\affiliation{Institute of Physics, National Yang Ming Chiao Tung University, Hsinchu, 300, Taiwan}

\author{Guey-Lin Lin}
\affiliation{Institute of Physics, National Yang Ming Chiao Tung University, Hsinchu, 300, Taiwan}

\begin{abstract}
Flavor transition mechanisms of supernova (SN) neutrinos during their propagation deserve a close scrutiny.  We present a method to verify Mikheyev-Smirnov-Wolfenstein (MSW) effect during the propagation of SN neutrinos from the SN core to the Earth. 
The non-MSW scenarios to be distinguished from the MSW one are the incoherent flavor transition probability for neutrino propagation in the vacuum and the flavor equalization
induced by fast flavor conversions.   Our approach involves studying the time evolution of neutrino event rates in liquid argon, liquid scintillation, and water Cherenkov detectors. The liquid argon detector is sensitive to $\nu_e$ flux while 
liquid scintillation and water Cherenkov detectors can measure $\bar{\nu}_e$ flux through inverse $\beta$ decay process.
The flux of $\nu_e$ ($\bar{\nu}_e$) is a linear combination of $\nu_e$ ($\bar{\nu}_e$) and $\nu_{\mu,\tau}$ ($\bar{\nu}_{\mu,\tau}$) fluxes from the source with the weighting of each component dictated by the flavor transition mechanism. 
%The former (latter) detector is sensitive to $\nu_e$ ($\bar{\nu}_e$) flux which is a linear combination of $\nu_e$ ($\bar{\nu}_e$) and $\nu_{\mu,\tau}$ ($\bar{\nu}_{\mu,\tau}$) fluxes from the source with the weighting of each component dictated by the flavor transition mechanism. 
Using currently available simulations for SN neutrino emissions, the time evolution of $\nu_e{\rm Ar}$ and $\bar{\nu}_e$ inverse $\beta$ decay event rates and the corresponding cumulative event fractions are calculated up to $t=100~{\rm ms}$ in DUNE, JUNO, 
and Hyper-Kamiokande detectors, respectively.  It is shown that the area under the cumulative time distribution curve from $t=0$ to $t=100~{\rm ms}$ in each detector and their ratio can be used to discriminate different flavor transition scenarios of SN neutrinos. 

\vspace{3mm}


\end{abstract}

\maketitle

\section{Introduction}

The flavor transition of supernova (SN) neutrinos has been an attractive field of research and motivated numerous efforts (see \cite{Mirizzi:2015eza} for a review) on studying the neutrino flavor conversions during the gravitational core collapse of a massive star. Originating from deep inside the SN core, neutrinos are expected to undergo significant flavor transitions as they propagate outward from a SN core to the terrestrial detectors. Studies have revealed that SN neutrino flavor conversions are induced when neutrinos experience significant matter potential, $\lambda=\sqrt{2}G_F n_e$, and/or $\nu-\nu$ potential, $\mu\sim\sqrt{2}G_F n_\nu$. Here $n_e$ and $n_\nu$ are electron and neutrino number densities, respectively, inside the SN.

Due to Mikheyev-Smirnov-Wolfenstein (MSW) effects~\cite{Wolfenstein:1977ue,Mikheev:1986gs}, $\nu_e$ ($\bar{\nu}_e$) flux swap with $\nu_{\mu,\tau}$ ($\bar{\nu}_{\mu,\tau}$) fluxes fully or partially when the neutrino vacuum oscillation frequency $\omega\equiv \Delta m^2/2E$ is of the order of the matter potential, $\omega\simeq\lambda$ \cite{Dighe:1999bi}. Here $\Delta m^2$ denotes one of the mass-squared differences and $E$ the neutrino energy.  

When neutrino densities are large, the off-diagonal $\nu-\nu$ potential, $\mu\sim\sqrt{2}G_F n_\nu$, arising from coherent $\nu-\nu$ forward scatterings, may induce collective pair flavor oscillation $\nu_e\bar{\nu}_e\leftrightarrow\nu_x\bar{\nu}_x$ with a frequency $\sim\sqrt{\omega\mu}$ over the entire energy range where $x=\mu,\tau$. Based on theoretical investigations and numerical calculations, large collective flavor conversions were predicted to occur 
when $\omega\simeq\mu$ \cite{Duan:2006an,Hannestad:2006nj,Duan:2010bg}. 

The $\nu-\nu$ potential may induce even faster flavor conversions at a rate $\sim\mu$ than the above collective oscillation at a rate $\sim\sqrt{\omega\mu}$ \cite{Sawyer:2005jk}. This fast flavor conversion requires sufficiently different angular distributions for different neutrino flavors and may lead to an equalization of neutrino fluxes of different flavors~\cite{Sawyer:2008zs,Sawyer:2015dsa,Chakraborty:2016lct}. This requirement can be fulfilled by different decoupling times of different flavors from matter.  
Since the flavor $\nu_x$~\footnote{Hereafter $\nu_x$ refers to $\nu_{\mu,\tau}$ and $\bar{\nu}_{\mu,\tau}$ for convenience in notation.} decouples from matter earlier than $\bar{\nu}_e$, and the latter decouples earlier than $\nu_e$, it can be expected that 
the $\nu_x$ zenith-angle distribution would be more forward peaked than that of $\bar{\nu}_e$, which in turn would be more forward peaked than the $\nu_e$ distribution. Therefore, the conditions required for fast flavor conversions are fulfilled.

Flavor transitions are expected to change flavor compositions of primary SN neutrino fluxes, and consequently to leave imprints on neutrino events measured by terrestrial detectors. This motivates us to study neutrino flavor transitions with measurements of galactic SN neutrinos arriving at the Earth. In different era during the SN explosion, different flavor transition scenario may dominate overs others. During the neutronization burst, the overwhelmingly huge flux of $\nu_e$ would suppress fast flavor conversions and leave only MSW effects to take their place~\cite{Hannestad:2006nj}. During the accretion phase, neutrinos are largely generated in all flavors with significant differences in flux spectra between electron and nonelectron flavors. 
These spectral differences shall lead to prominent flavor transition effects in the accretion phase, which may be solely due to MSW or with the effect of fast flavor conversions as well~\cite{Capozzi:2018rzl}.
In the final cooling phase, the neutrino spectrum of each flavor is quite similar to each other so that the flavor transition effects are not significant. 

We note that the discrimination between flavor transitions due to MSW effects and the flavor equalization (FE) from fast flavor conversions has been well studied in the accretion phase~\cite{Capozzi:2018rzl}. Here we shall focus on MSW effects to the propagation of SN neutrinos in the era of a neutronization burst. Based on the understanding that MSW effects are sensitive to the neutrino mass ordering, many studies~\cite{Lunardini:2003eh,Dasgupta:2008my,Duan:2007bt,Serpico:2011ir,Chiu:2013dya,Lai:2016yvu,Lee:2018kup,Vale:2015pca} are devoted to probing such an ordering with SN neutrino events detected on the Earth. Naturally, these studies all assume the occurrence of MSW effects. Although it has been well understood that the MSW effect happens in
the propagation of solar neutrinos, 
such effects on the propagation of SN neutrinos are far more nontrivial in the following two aspects. First, such effects are operative for both neutrinos and antineutrinos. Second, they are sensitive to the value of the neutrino mixing parameter $|U_{e3}|^2$~\cite{Dighe:1999bi}. Recent measurements of this parameter are summarized in Ref.~\cite{Workman:2022ynf} where original references are given. Results of these measurements imply that 
MSW flavor conversions of SN neutrinos are in the adiabatic regime.
Due to these intriguing properties, it is worthwhile to verify whether MSW effects really occur or not in the era of a neutronization burst.  
To do this, we compute time dependencies of SN neutrino event rates predicted by MSW effects for normal ordering (NO) and inverted ordering (IO), respectively. The first non-MSW scenario to be distinguished from the MSW one is the vacuum oscillation, which reduces to incoherent flavor transition probability~\cite{Learned:1994wg,Athar:2000yw,Bento:1999bb} for neutrinos traversing a vast distance. Time dependencies of SN neutrino event rates in this scenario, which will be referred to as a vacuum flavor transition (VFT) hereafter, are also calculated. 
The second non-MSW scenario to be compared with MSW one is the FE just mentioned. Although this scenario were argued to be important only in the accretion phase, it is of interest to 
directly verify its role in the neutronization era with the corresponding time dependencies of SN neutrino event rates.     
In NO, it is to be seen that the $\nu_e$ event rate is sufficient to isolate MSW from VFT and FE. However, for IO, one needs to invoke both $\nu_e$ and $\bar{\nu}_e$ event rates for distinguishing various flavor transition mechanisms.   

This paper is organized as follows. In Sec. II, we briefly review possible flavor transition mechanisms of SN neutrinos as they propagate outward from SN core until reaching the terrestrial detector. We then summarize SN neutrino fluxes obtained from the simulated SN neutrino data, which will be used in our later analysis. In Sec. III, we calculate event rates of $\nu_e{\rm Ar}$ interactions for DUNE detector and $\bar{\nu}_e$ inverse $\beta$ decay (IBD) event rates for JUNO and Hyper-Kamiokande (HyperK) detectors under different flavor transition mechanisms, taking a galactic SN burst at $5$ kpc distance as a benchmark SN neutrino source.  In Sec. IV, we discuss the strategy of testing MSW effects with $\nu_e{\rm Ar}$ and $\bar{\nu}_e$ IBD event rates. We first present cumulative time distributions of neutrino events expected in DUNE detector for $\nu_e{\rm Ar}$ interactions and JUNO detector representing for $\bar{\nu}_e$ IBD interactions. Since cumulative time distributions of HyperK events are similar to those of JUNO events (except on statistical uncertainties), we do not present them here but incorporate them in the latter analysis. We next integrate the above cumulative time distributions over our interested time range $0\leq t \leq 100$ ms. Values of these integrals are effective for discriminating different flavor transition scenarios. Specifically, for the benchmark case of $5$ kpc far SN, we shall see that integrals over cumulative time distributions of $\nu_e{\rm Ar}$ events in different flavor transition scenarios are sufficient to separate MSW-NO from MSW-IO, FE, and VFT. On the other hand, to discriminate between the latter three scenarios, it is necessary to invoke integrals arising from both $\nu_e{\rm Ar}$ and $\bar{\nu}_e$ IBD event distributions. For the latter type of integrals, we include expected results from JUNO and HyperK detectors. The effect of SN distance to the discriminating power of our method will also be presented. Finally we also briefly discuss Earth matter effects to the above integrals. It will be shown that Earth matter effects are negligible in our analysis.       
We summarize and conclude in Sec. V. 

\section{Supernova Neutrino Flux Spectra}

\subsection{Primary neutrino flux spectra}\label{NuFlux}

A SN neutrino burst lasts for $\Delta t\approx10~{\rm s}$, during which the neutronization burst happens at $t_{pb}\sim (10-15)~{\rm ms}$. Here, $t_{pb}$ denotes the postbounce time. In our calculation, the primary neutrino flux spectra are extracted from SN simulations accounting for SNe with iron core. Simulations of SN explosion have been pursued by different groups.  
To demonstrate our approach, we calculate expected SN neutrino event rates based upon neutrino emissions simulated by four different groups with respect to roughly similar progenitor masses. These simulations are for progenitor masses of $8.8~{\rm M}_\odot$ by Garching group \cite{Huedepohl:2009wh}, of $10~{\rm M}_\odot$ by Burrow {\it et al.} \cite{Burrows:2019rtd}, of $11.2~{\rm M}_\odot$ by Fischer {\it et al.} \cite{Fischer:2015sll}, and of $13~{\rm M}_\odot$ by Nakazato {\it et al.}~\cite{Nakazato:2012qf}. We note that the simulations in Refs.~\cite{Burrows:2019rtd,Nakazato:2012qf} cover many different progenitor masses. The above specific choices of progenitor masses are made for matching with studies of Refs.~\cite{Huedepohl:2009wh,Fischer:2015sll}.
Finally, we take the distance between the galactic SN and the Earth as $5$ kpc for a benchmark discussion.  

\begin{figure}[htbp]
	\begin{center}
	\includegraphics[width=.5\columnwidth]{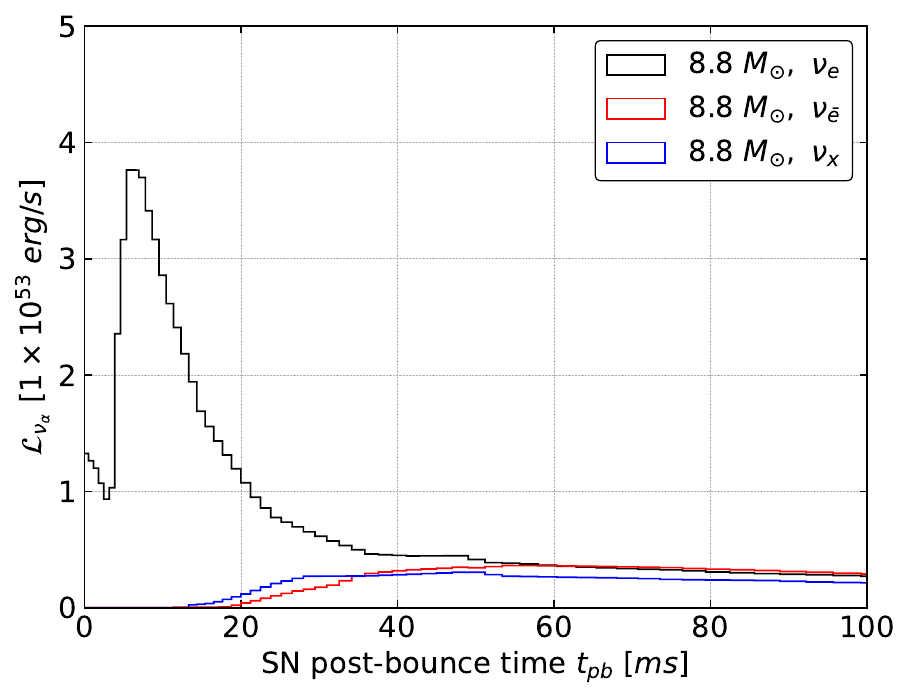}%
	\includegraphics[width=.5\columnwidth]{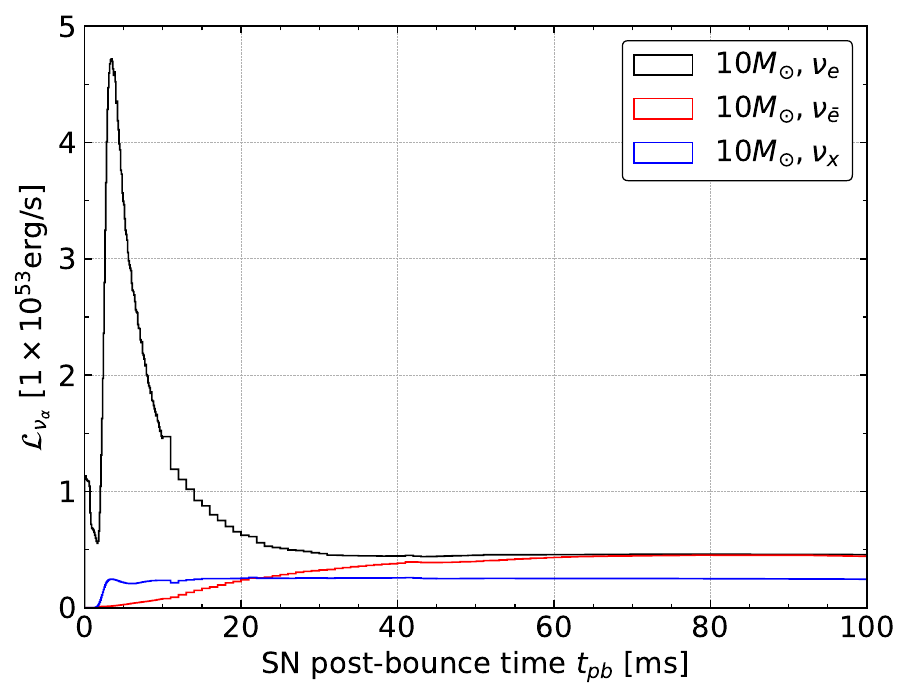}\\
	\includegraphics[width=.5\columnwidth]{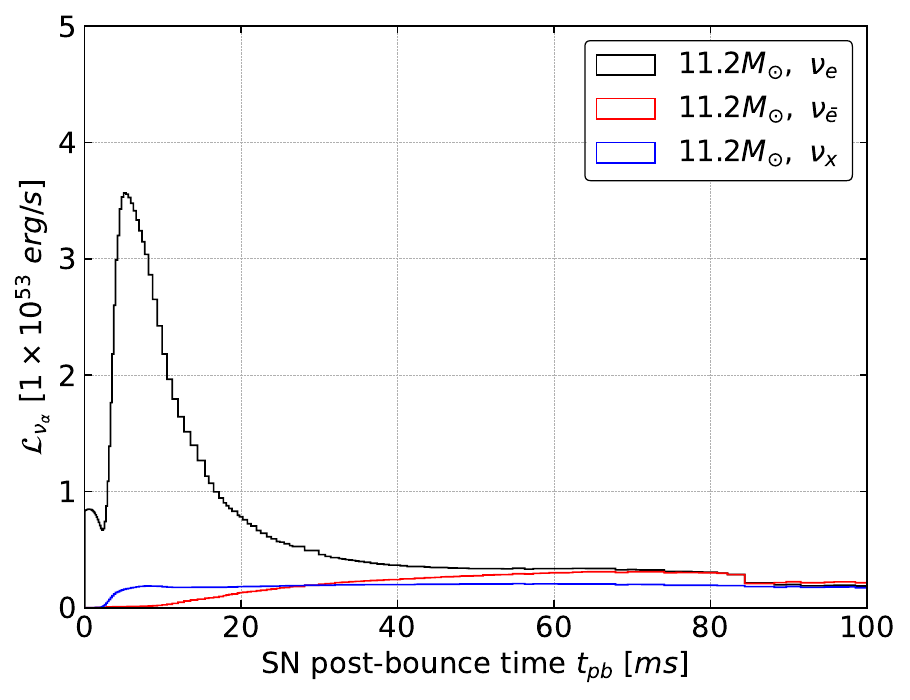}%
	\includegraphics[width=.5\columnwidth]{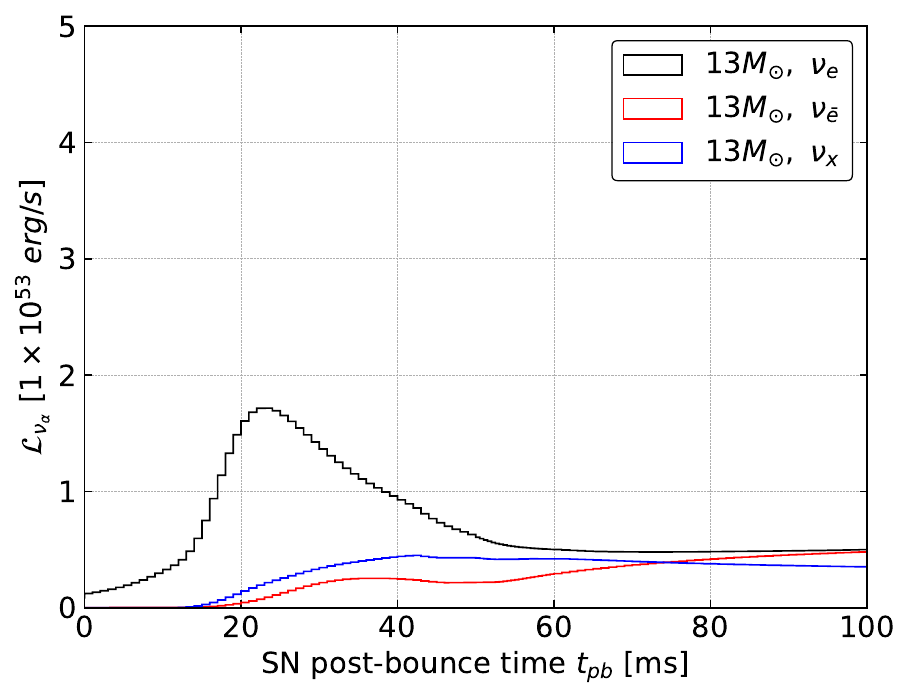}%
	\caption{Luminosities ${\mathcal L}_{\nu_\alpha}$ predicted by simulations G (upper left), B (upper right), F (bottom left), and N (bottom right) for progenitor masses of $8.8$, $10$, $11.2$, and $13~{\rm M}_\odot$, respectively. Here $\alpha=\{e,\bar{e},x\}$ is the flavor index, and $\left< E_{\nu_\alpha} \right>$ is the average energy of neutrinos of flavor $\alpha$.} 
	\label{luminosity}
	\end{center}
\end{figure}

\begin{figure}[htbp]
	\begin{center}
		\includegraphics[width=.5\columnwidth]{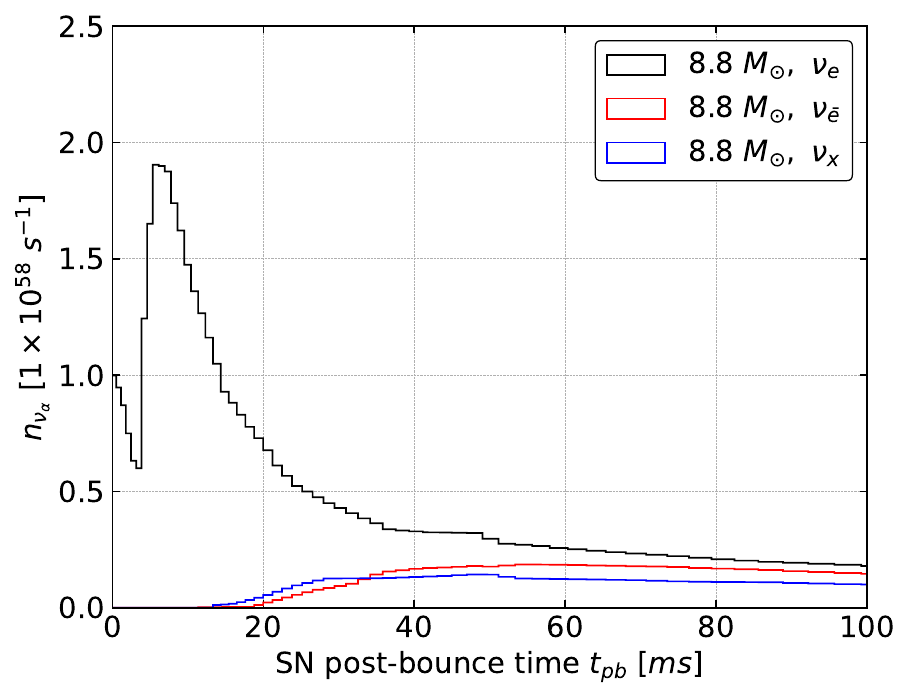}%
		\includegraphics[width=.5\columnwidth]{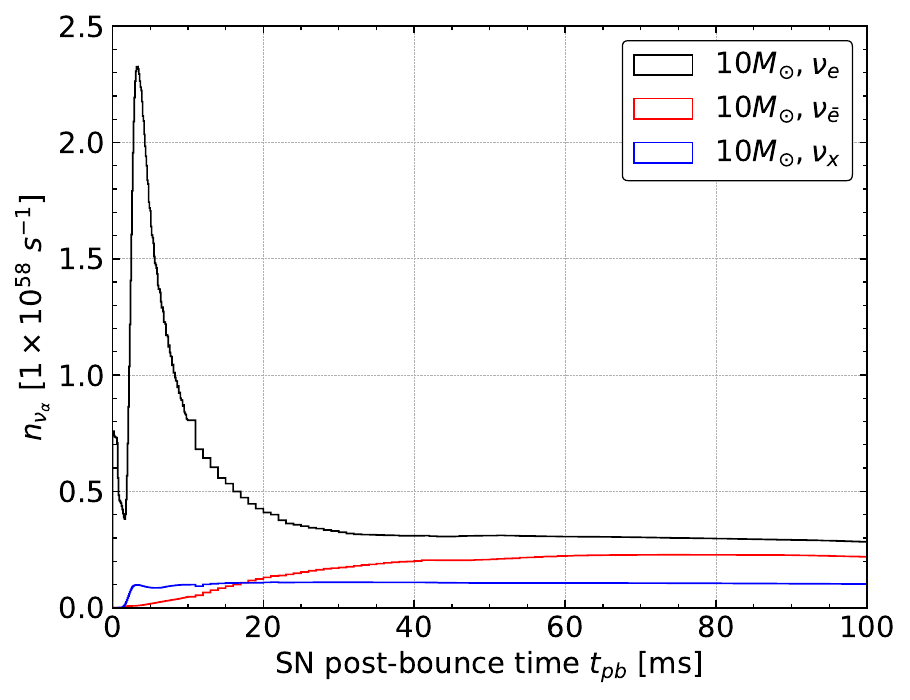}\\
		\includegraphics[width=.5\columnwidth]{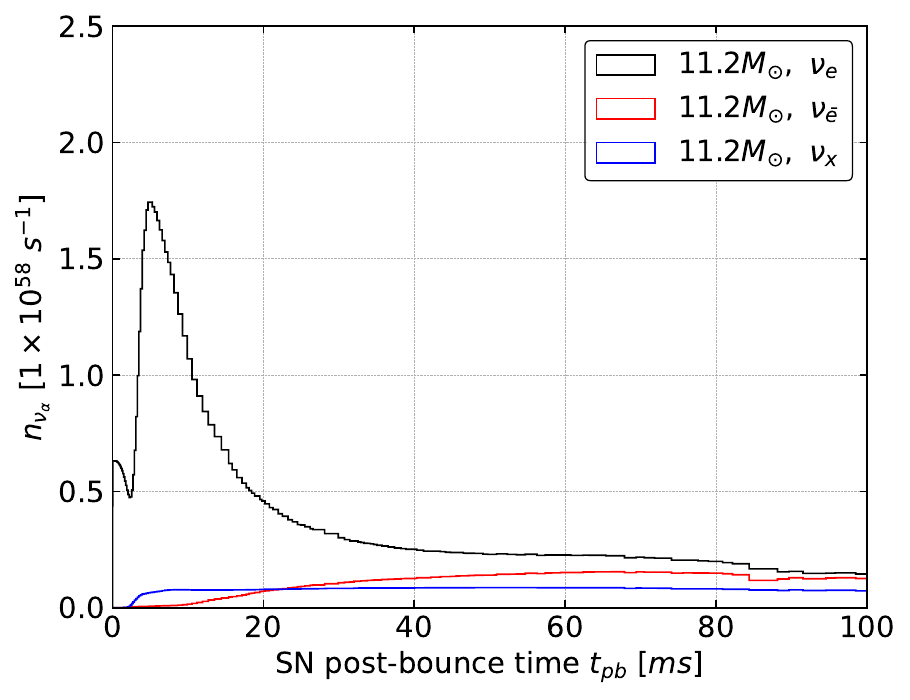}%
		\includegraphics[width=.5\columnwidth]{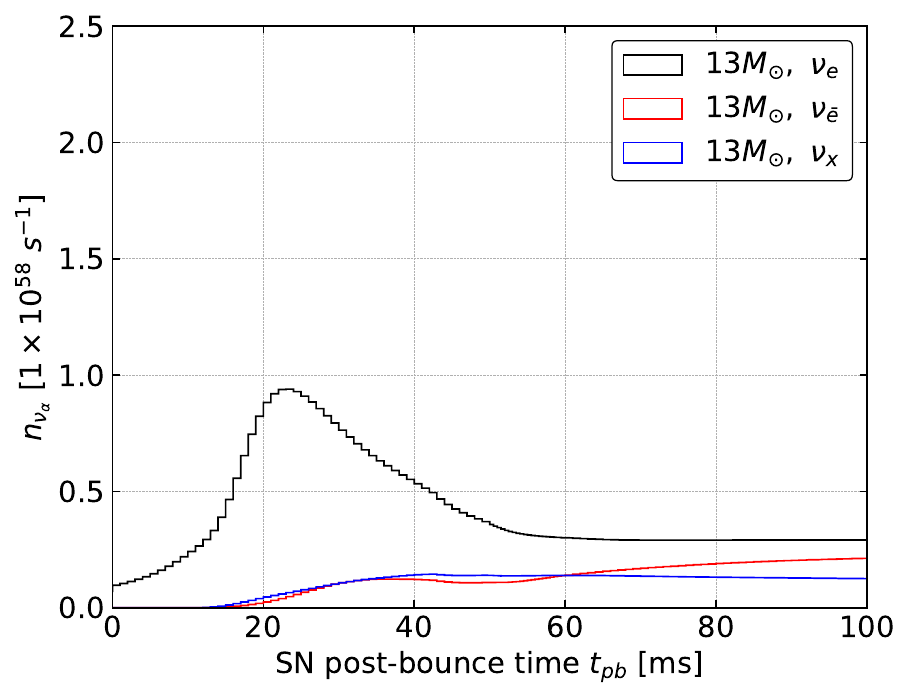}%
		\caption{Emission rates $n_{\nu_\alpha}\equiv {\mathcal L}_{\nu_\alpha}/ \left< E_{\nu_\alpha} \right>$ predicted by simulations G, B, F, and N (arranged in the same way as Fig.~\ref{luminosity}) for progenitor masses of $8.8$, $10$, $11.2$, and $13~{\rm M}_\odot$, respectively.} 
		\label{emsrate} %this is a new label!!!!!!!!!!!!!
	\end{center}
\end{figure}

The SN luminosity $\mathcal{L}_{\nu_\alpha}$ (erg/s) and emission rate $n_{\nu_\alpha}$ (1/s) of neutrinos $\nu_{\alpha}$ are shown in Figs.~\ref{luminosity} and \ref{emsrate}, respectively. In each figure, the upper left is the result  predicted by Garching simulation (simulation G), upper right is by Burrow {\it et al.}'s simulation (simulation B), lower right is by Fischer {\it et al.}'s simulation (simulation F), and lower right is by Nakazato {\it et al.}'s simulation (simulation N). Clearly, the luminosity and emission curves predicted by simulations G, B, and F have similar time dependencies while those of simulation N are quite different from the above. 
In simulation N with respect to $\mathcal{L}_{\nu_\alpha}$, the neutronization burst occurs at $t_{pb}\sim10~{\rm ms}$ with its full width at half maximum $\Delta t_N\sim30~{\rm ms}$ while, in the other simulations, the neutronization burst occurs earlier with $\Delta t_N\sim10~{\rm ms}$. One also observes that simulation N predicts a peak luminosity twice larger than the tail one while any of the other simulations predicts a peak luminosity 10 times larger than the tail one. To cover the whole duration of the neutronization burst, we perform the analyses in a time period of $\Delta t=100~{\rm ms}$ from $t_{pb}=0$.

The primary SN neutrino flux spectra of flavor $\alpha$ without any flavor transitions during the neutrino propagation
are well fitted by the Keil parametrization~\cite{Keil:2002in}, i.e., 
\begin{equation}
F^0_\alpha(E,t)=\frac{1}{4\pi d^2}\left(\frac{d^2N_{\nu_\alpha}}{dt dE}\right),   
\end{equation}
with~\footnote{Here we adopt the notation $F^0_{\alpha}$ instead of $F^0_{\nu_\alpha}$ to denote the flux spectra of $\nu_\alpha$. This notation simplifies Eq.~(\ref{tran_p}).}
\begin{equation}
 \frac{d^2N_{\nu_\alpha}}{dt dE}= \frac{n_{\nu_\alpha}}{ \left< E_{\nu_\alpha} \right>}\frac{(1+\eta_{\nu_\alpha})^{(1+\eta_{\nu_\alpha})}}{\Gamma(1+\eta_{\nu_\alpha})}\left(\frac{E}{\left< E_{\nu_\alpha} \right> }\right)^{\eta_{\nu_\alpha}} 
                          \exp\left[-(\eta_{\nu_\alpha}+1)\frac{E}{\left< E_{\nu_\alpha}\right> }\right],
\label{flux_spectra}
\end{equation}
where 
$\eta_{\nu_\alpha}$ specifies the pinching of the spectrum, and $d$ is the distance between SN and the Earth. For our interested time period, $0\leq t\leq 100~{\rm ms}$, the relevant parameters in Eq.~(\ref{flux_spectra}) is summarized in Table \ref{emission}. The mean energies, $(\langle E_{\nu_e}\rangle,~\langle E_{\bar{\nu}_e}\rangle,~\langle E_{\nu_x}\rangle)$, in simulation G are the most degenerate while those in simulation N are the most diverse. While the $\langle E_{\nu_e}\rangle$s are about $10~{\rm MeV}$ in all four simulations, $\langle E_{\nu_x}\rangle$ ranges from $9.9~{\rm MeV}$ in simulation G to $18.7~{\rm MeV}$ in simulation N.

\begin{table}[htbp]
\begin{center}
\begin{tabular}{lrrrrrrrrrrrr}\hline\hline
 Simulations                & \multicolumn{3}{c}{G} & \multicolumn{3}{c}{B} & \multicolumn{3}{c}{F} & \multicolumn{3}{c}{N}  \\ \cline{2-13}
 Flavor      &  $\nu_e$  &  $\bar{\nu}_e$  &  $\nu_x$  &  $\nu_e$  &  $\bar{\nu}_e$  &  $\nu_x$  &  $\nu_e$  &  $\bar{\nu}_e$    & $\nu_x$ &   $\nu_e$  &  $\bar{\nu}_e$  & $\nu_x$ \\ \hline
 $\langle E_{\nu_\alpha}\rangle$ [MeV]  &  ~9.3    &  ~9.1   &   ~9.9    &  10.3   &   12.1    &    14.7    &   10.6   &   11.9   &  14.7    &   10.5   &  13.4   &    18.7 \\ 
${\mathcal E}_{\nu_\alpha}~[10^{51}~{\rm erg}]$    &  9.1  &   2.6   &   2.2   &   7.2   &   3.5   &   2.5   &   6.4   &   2.1  &   1.9    &   7.4   &   2.7   &   3.2   \\
 ${\mathcal N}_{\nu_\alpha}~[10^{56}]$       &   5.7   &  1.3   &  1.0   &  4.4  &  1.8   &   1.0   &   3.8  &   1.1   &  0.8       &   4.4   &   1.3   &   1.1  \\  
     $\eta_{\nu_\alpha}$       &  3.8    &   3.0   &  2.2  &  5.2   &   5.1  &   4.1  &  4.4   &  4.4   &  2.5     &   3.6   &   2.1   &   1.8 \\   \hline

\end{tabular}
\end{center}
\caption{Mean energy $\langle E_{\nu_\alpha}\rangle$, energy emission ${\mathcal E}_{\nu_\alpha}\equiv \int dt \, {\mathcal L}_{\nu_\alpha}$, total number emitted $\mathcal{N}_{\nu_\alpha}\equiv \int dt \, n_{\nu_\alpha}$, and pinching $\eta_{\nu_\alpha}$ in Keil parametrization of SN neutrino flux spectra given by Eq.~(\ref{flux_spectra}).}
\label{emission}
\end{table}

\label{flux}

\subsection{Neutrino flux spectra with flavor transitions}\label{nuEarth}

In this paper, we investigate whether or not the flavor contents of SN neutrinos are modified by the MSW effect as they propagate outwards from deep inside a SN and finally reaches  the Earth. If MSW effects do not occur, the flavor contents of SN neutrinos arriving on the Earth are the incoherent superposition of mass eigenstates leaving from SN, which can be written as
\begin{equation}
F_\alpha(E,t) = P_{\alpha\beta} F^0_\beta(E,t),
\label{tran_p}
\end{equation}
where the flavor transition probability $P_{\alpha\beta}\equiv P(\nu_\beta\rightarrow\nu_\alpha)$ is given by~\cite{Learned:1994wg,Athar:2000yw,Bento:1999bb}  
\begin{equation}
P_{\alpha\beta} = \sum_k |U_{\alpha k}|^2 |U_{\beta k}|^2.
\label{incoherent}
\end{equation}
Here $U$ is PMNS mixing matrix of neutrinos~\cite{PMNS} and the flavor $\alpha$ runs for both neutrinos and antineutrinos. In terms of mixing angles, $U$ is given by
\begin{equation}
U = \left(
                            \begin{array}{ccc}
                            1 &             0               & 0 \\
                            0 &  \cos\theta_{23}    &  \sin\theta_{23} \\
                            0 &  -\sin\theta_{23}    &  \cos\theta_{23}
                            \end{array}
                            \right)
         \left(
                            \begin{array}{ccc}
                            \cos\theta_{13}                    &  0  &  \sin\theta_{13}e^{i\delta} \\
                            0                                          &  1  &  0  \\
                            -\sin\theta_{13}e^{-i\delta}   &  0  &  \cos\theta_{13}
                            \end{array}
                            \right) 
       \left(
                            \begin{array}{ccc}
                            \cos\theta_{12}    &    \sin\theta_{12}      &  0 \\
                            -\sin\theta_{12}    &   \cos\theta_{12}      &  0 \\
                            0                          &               0                &  1
                            \end{array}
                            \right),
\end{equation}
where the updated values for the mixing angles can be found from~\cite{Workman:2022ynf}.  
This scenario is the same as astrophysical neutrinos traversing a long distance in vacuum before reaching the Earth.  
As stated earlier, we have referred this scenario as a VFT. We stress again that Eq.~(\ref{incoherent}) is valid for both neutrinos and antineutrinos.

Using best-fit values of neutrino mixing angles~\cite{Esteban:2018azc,NMH_values} with the analysis of Super-Kamiokande atmospheric neutrino data included~\cite{Super-Kamiokande:2017yvm} for NO scenario, we have 
\begin{eqnarray}
P_{\alpha\beta}=
\left(
                            \begin{array}{ccc}
                            0.55 &             0.18               & 0.27 \\
                            0.18 &  0.44    &  0.38 \\
                            0.27 &  0.38    &  0.35
                            \end{array}
                            \right).    
\end{eqnarray}
Hence Eq.~(\ref{tran_p}) becomes 
\begin{eqnarray}
F_e             & = &  0.55 F_e^0 + 0.45 F_x^0, \label{eVO}  \\
F_{\bar{e}}  & = &  0.55 F_{\bar{e}}^0 + 0.45F_x^0,    \label{ebarVO}  \\
4F_x           & = &  0.45 F_e^0 + 0.45 F_{\bar{e}}^0 + 3.10 F_x^0,    \label{xVO}
\end{eqnarray}
where $4F_x \equiv F_\mu + F_{\bar{\mu}}+F_\tau+F_{\bar{\tau}}$.
Best-fit values of mixing angles in IO do not give noticeable changes on the above equations nor do mixing angles fitted by other groups~\cite{Capozzi:2018ubv,deSalas:2017kay} mentioned in~\cite{Workman:2022ynf}.  

In MSW scenarios, the flux spectra arriving at the detector on Earth are given by 
\begin{eqnarray} 
F_e            & = &  F^0_x,   \label{eNH}  \\
F_{\bar{e}} & = &  (1-\bar{P}_{2e}) F^0_{\bar{e}} + \bar{P}_{2e} F^0_{\bar{x}}, \label{ebarNH}  \\
4F_x          & = & F^0_e + F^0_{\bar{e}} + 4F^0_x - F_e - F_{\bar{e}} = F^0_e + \bar{P}_{2e} F^0_{\bar{e}} + (3-\bar{P}_{2e}) F^0_x,  \label{xNH}
\end{eqnarray}
for the NO, and
\begin{eqnarray} 
F_e            & = & P_{2e} F^0_e + (1-P_{2e}) F^0_x,  \label{eIH} \\
F_{\bar{e}} & = & F^0_{\bar{x}}, \label{ebarIH}  \\
4F_x          & = & F^0_e + F^0_{\bar{e}} + 4F^0_x - F_e - F_{\bar{e}} = (1-P_{2e}) F^0_e + F^0_{\bar{e}} + (2+P_{2e}) F^0_x,  \label{xIH}
\end{eqnarray}
for IO \cite{Dighe:1999bi}. Here $P_{2e}$ ($\bar{P}_{2e}$) is the probability that a mass eigenstate $\nu_2$ ($\bar{\nu}_2$) is observed as $\nu_e$ ($\bar{\nu}_e$) when it reaches the terrestrial detector. 
Without taking into account Earth matter effects, $P_{2e}= |U_{e2}|^2=\sin^2\theta_{12}+{\mathcal O}(\sin^2\theta_{13})$. 
We can simply take $P_{2e}=\sin^2\theta_{12}$ by disregarding ${\mathcal O}(\sin^2\theta_{13})$ contributions.   
In fact, contributions of the same order are also neglected from Eqs.~(\ref{eNH}) to (\ref{xIH}).
The best-fit value for $\sin^2\theta_{12}$ is $0.310$ for both NO and IO~\cite{Esteban:2018azc}.  

In MSW scenarios, one can see that $\nu_e$ completely comes from $\nu_x^{0}$ from the source while $\bar{\nu}_e$ comes from both $\bar{\nu}_e^{0}$ and $\bar{\nu}_x^{0}$ for NO.  
On the other hand, for IO, $\nu_e$ comes from both $\nu_e^{0}$ and $\nu_x^{0}$ while $\bar{\nu}_e$ completely comes from $\bar{\nu}_x^{0}$. 

Finally for the FE scenario, we have 
\begin{eqnarray} 
F_e =  F_{\bar{e}}=F_x=\frac{1}{3}\left(F_e^0+F_{\bar{e}}^0+2F_x^0\right).          
 \label{FE}
\end{eqnarray}

Before moving on, we reiterate that Earth matter effects have not been taken into account so far. Such effects will be estimated in Sec. IV and and shown to be negligible in our analysis.

\section{Event Rates of SN neutrinos in Terrestrial Detectors %Around The Neutronization Burst
}

With neutrino fluxes given above, we calculate event rates of SN neutrinos for all flavors, $\nu_e$, $\bar{\nu}_e$, and $\nu_x$, for VFT scenario, the scenario that the flavor contents are modified by MSW effects as SN neutrinos propagate outward from the core, and finally the FE scenario induced by 
the fast flavor conversions.  In the second case, both NO and IO scenarios are taken into consideration and denoted as MSW-NO and MSW-IO, respectively. The event rates and quantities induced from these rates are displayed in numbers per bin with a $5~{\rm ms}$ bin width throughout this article.

\begin{figure}[htbp]
	\begin{center}
	\includegraphics[width=.5\columnwidth]{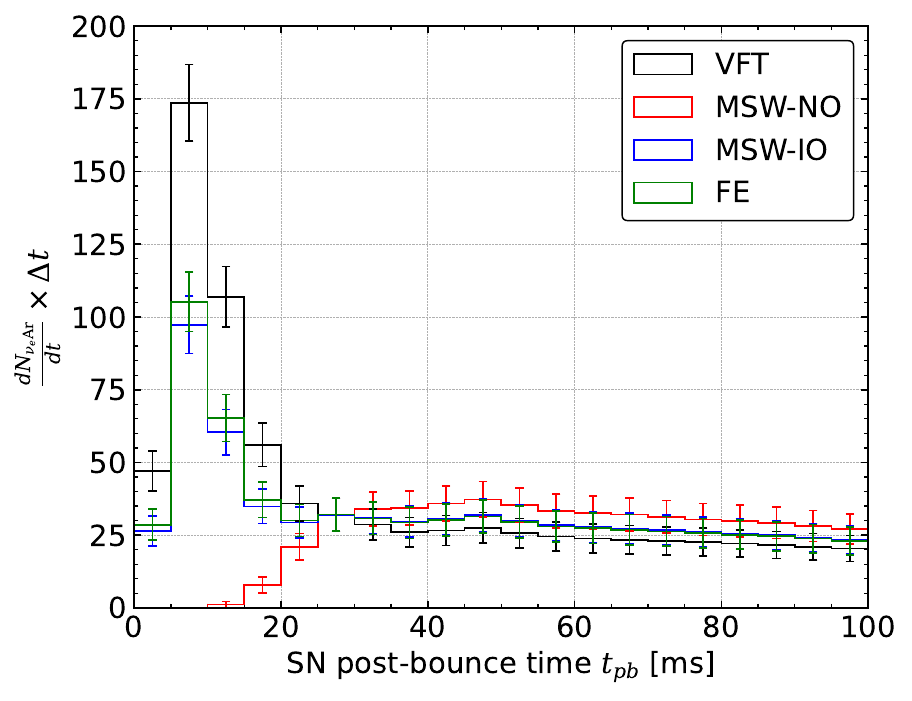}%
	\includegraphics[width=.5\columnwidth]{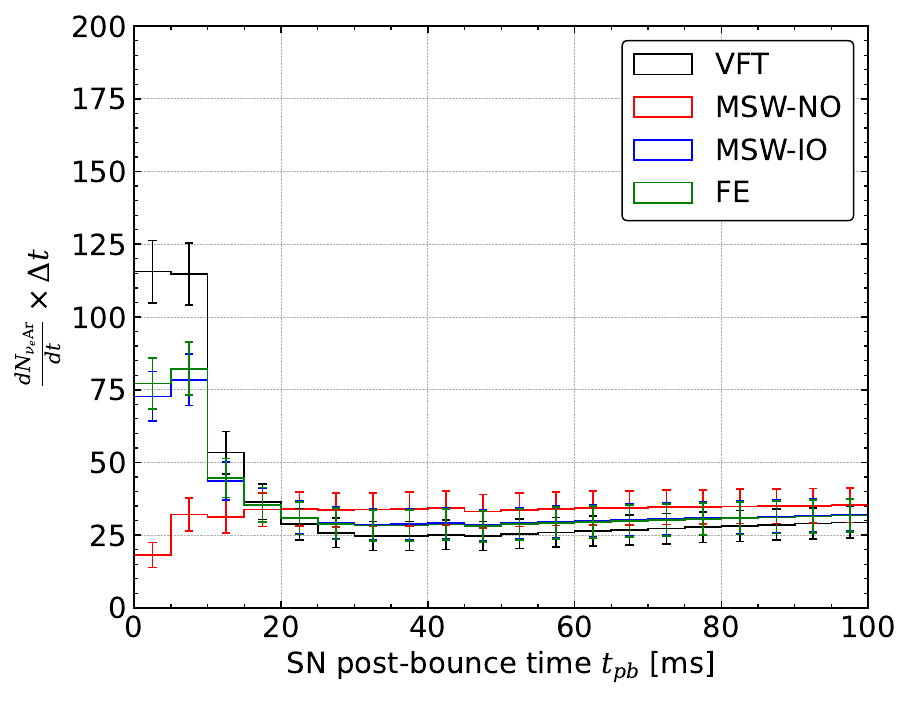}\\
	\includegraphics[width=.5\columnwidth]{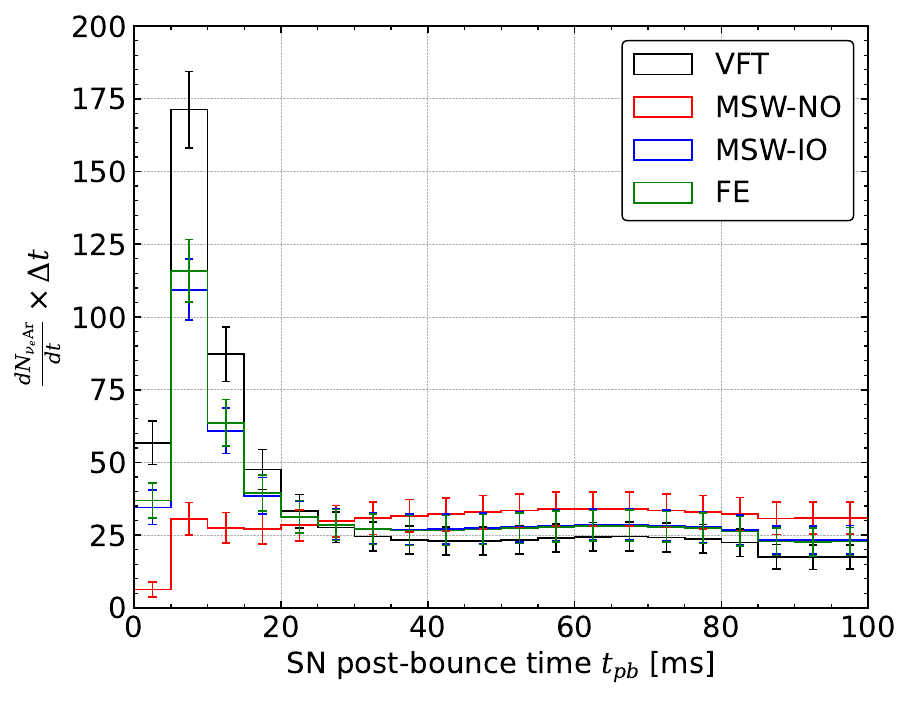}%
	\includegraphics[width=.5\columnwidth]{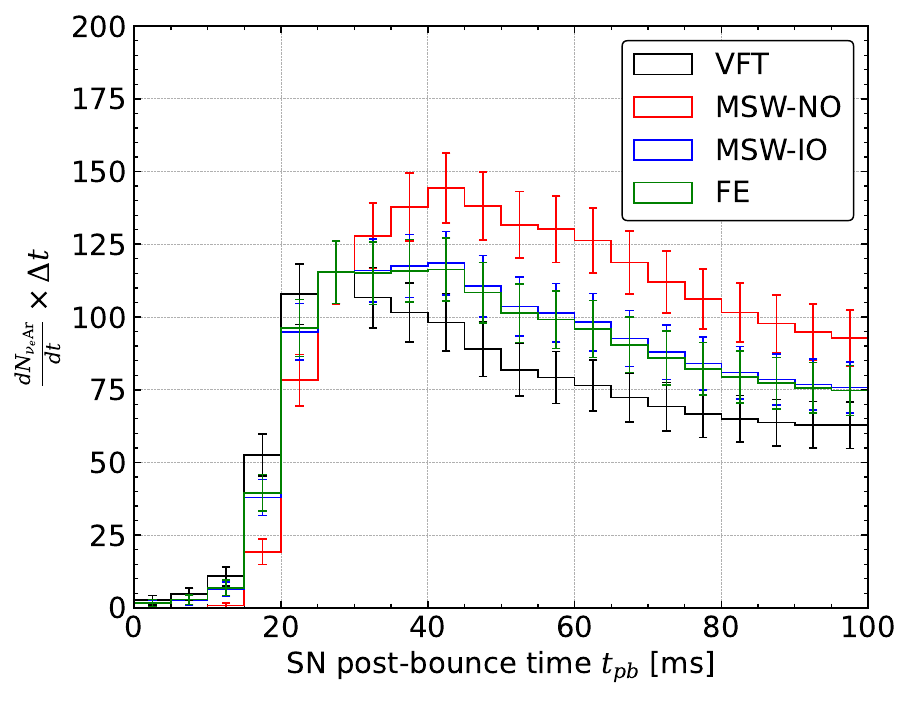}%
	\caption{Event rates of $\nu_e{\rm Ar}$ in DUNE for different flavor transition scenarios. These event rates are obtained for simulations G, B, F, and N (arranged in the same way as Fig.~\ref{luminosity}) with progenitor masses of $8.8$, $10$, $11.2$, and $13~{\rm M}_\odot$, respectively.}
	\label{duneAr}
	\end{center}
\end{figure}

In liquid argon time projection chambers, $\nu_e$ is the most easily detected species via its charged-current interaction with argon nuclei, $\nu_e+{^{40}{\rm Ar}}\rightarrow{^{40}{\rm K}^*}+e^-$. The cross section for this $\nu_e{\rm Ar}$ 
interaction has been computed in~\cite{Kolbe:2003ys}. Numerical data compiled in~\cite{Scholberg:2012id} is used for our subsequent analyses. Assuming a SN at a distance of $5~{\rm kpc}$, the event spectrum of $\nu_e{\rm Ar}$ in DUNE~\cite{DUNE:2015lol} is given by
\begin{eqnarray}
	&& \left(\frac{d^2N_{\nu_e{\rm Ar}}}{dE_{e^-}dt}\right)\Delta t=N_{\rm Ar}\cdot\int dE_\nu F_e(E_{\nu},t)\Delta t\cdot\frac{d\sigma_{\nu_e{\rm Ar}}(E_\nu,~E_{e^-})}{dE_{e^-}}, \label{Arspec} 
\end{eqnarray}
where $\Delta t\equiv 5$ ms is our chosen bin width, $N_{\rm Ar}$ is the number of target liquid argon in DUNE detector. Integrating the electron energy, we obtain $\nu_e {\rm Ar}$ number $(dN_{\nu_e{\rm Ar}}/dt)\Delta t$ per $5$ ms as  
shown in Fig. \ref{duneAr}. It is clearly seen that the time-dependence profiles of the event rates for simulations G, B, and F are similar while the profile for simulation N is rather different from the others.

For simulations G, B, and F,
the peak event rates of VFT are larger than those in FE while the peak event rates of FE are larger than those in MSW, in which the peak event rates of IO are larger than those of NO. Hence the ordering of peak event rates is VFT$>$FE$>$MSW-IO$>$MSW-NO. This can be understood from peak behaviors of $\mathcal{L}_{\nu_\alpha}$ and $n_{\nu_\alpha}$ shown in Figs.~\ref{luminosity} and \ref{emsrate}, respectively. We can see that $\mathcal{L}_{\nu_e} (n_{\nu_e})\gg \mathcal{L}_{\nu_x} (n_{\nu_x})$ in the peak region so that $F_e$ is approximately equal to $0.55F_e^0$, $0.33F_e^0$, $0.31F_e^0$, and $F_x^0$ for VFT, FE, MSW-IO and MSW-NO, respectively. Such an ordering for peak $\nu_e$ flux is preserved in the event rate level since the difference between $\langle E_{\nu_e}\rangle$ and $\langle E_{\nu_x}\rangle$ is not sufficient to flip the ordering. 
For the long tail region, $F_e^0$ and $F_x^0$ are both small and comparable to each other so that, after their full or partial swap, the resulting event rates in different scenarios are indistinguishable with uncertainties taken into account.

For simulation N, the event rates in all scenarios  
show similar behaviors except on the magnitude and timing of the peak. The peak of VFT appears earlier than those of MSW and FE scenarios. On the other hand, the peak event rate of MSW-NO scenario is the largest. 
Here the ordering of peak event rates are different because 
$\mathcal{L}_{\nu_x} (n_{\nu_x})$ is not so much smaller than $\mathcal{L}_{\nu_e} (n_{\nu_e})$ as in the case of simulations G, B, and F. Furthermore the contribution of $F_x^0$ to $\nu_e$ event rate through the flavor transition is enhanced due to a relatively large $\langle E_{\nu_x}\rangle$ in simulation N. For the tail region, simulation N also predicts roughly similar $F_e$ for different flavor transition scenarios. However, due to large $\langle E_{\nu_x}\rangle$, the tail event rates for different scenarios 
are also determined by the fractions of $F_x^0$ in $F_e$. These fractions are given by Eqs.~(\ref{eVO}), (\ref{eNH}), (\ref{eIH}), and (\ref{FE}) which predict the ordering of event rates shown on the lower right of Fig.~\ref{duneAr}.     
It is seen that these event rates are much larger than those in simulations G, B, and F, which is again caused by the large $\langle E_{\nu_x}\rangle$ that enhances the event rates. 

%
%\begin{table}[htbp]
%	\begin{center}
%		\begin{tabular}{lcccccccccccc}\hline\hline
%			Model               & \multicolumn{3}{c}{G} & \multicolumn{3}{c}{B} & \multicolumn{3}{c}{F}  & \multicolumn{3}{c}{N}  \\ 
%			\cline{2-13}
%			%\cmidrule(lr){2-4} \cmidrule(lr){5-7} \cmidrule(lr){8-10} \cmidrule(lr){11-13}   
%			Signature  &  $\nu_e{\rm Ar}$  & JUNO & HyperK  &  $\nu_e{\rm Ar}$  & JUNO & HyperK &  $\nu_e{\rm Ar}$  &  JUNO & HyperK &   $\nu_e{\rm Ar}$  &  JUNO & HyperK \\ \hline
%			VFT  & 788 & 849 & 14404 & 751 & 985 & 16737 & 738 & 690 & 11719 & 1561& 1193 & 24829 \\ 
%			MSW-NO    & 585 & 854 & 14477 & 671 & 1000 & 16988 & 615 & 675 & 11449 & 2144 & 1088 & 23073 \\
%			MSW-IO       & 673 & 831 & 14122 & 716 & 925 & 15767 & 685 & 750 & 12768 & 1815 & 1600 &  31642 \\ 
%			FE       & 685 & 842 & 14293 & 719 & 961 & 16355 & 690 & 714 & 12132 & 1791 & 1614 &  27514 \\ 
%			\hline\hline
%			
%		\end{tabular}
%	\end{center}
%	\caption{Total numbers of SN neutrino events for $0\leq t^{*}\leq 100$ ms for $\nu_e{\rm Ar}$ by DUNE detector and IBD signals by JUNO and HyperK detectors in different flavor transition scenarios for simulations G, B, F, and N with progenitor masses of $8.8~{\rm M}_\odot$, $10~{\rm M}_\odot$, $11.2~{\rm M}_\odot$, and $13~{\rm M}_\odot$, respectively. }
%	\label{totnum}
%\end{table}

\begin{table}[htbp]
	\begin{center}
		\begin{tabular}{lcccccccccccc}\hline\hline
			Model               & \multicolumn{3}{c}{G} & \multicolumn{3}{c}{B} & \multicolumn{3}{c}{F}  & \multicolumn{3}{c}{N}  \\ 
			\cline{2-13}
			%\cmidrule(lr){2-4} \cmidrule(lr){5-7} \cmidrule(lr){8-10} \cmidrule(lr){11-13}   
			Signature  &  $\nu_e{\rm Ar}$  & JUNO & HyperK  &  $\nu_e{\rm Ar}$  & JUNO & HyperK &  $\nu_e{\rm Ar}$  &  JUNO & HyperK &   $\nu_e{\rm Ar}$  &  JUNO & HyperK \\ \hline
			VFT  & 789 & 735 & 12474 & 749 & 973 & 16536 & 737 & 679 & 11527 & 1390& 1193 & 20323 \\ 
			MSW-NO    & 512 & 737 & 12495 & 664 & 988 & 16775 & 603 & 664 & 11263 & 1874 & 1088 & 18523 \\
			MSW-IO       & 668 & 729 & 12394 & 712 & 916 & 15605 & 678 & 738 & 12553 & 1601 & 1600 &  27308 \\ 
			FE       & 681 & 733 & 12442 & 716 & 950 & 16168 & 685 & 702 & 11932 & 1580 & 1353 &  23077 \\ 
			\hline\hline
			
		\end{tabular}
	\end{center}
	\caption{Total number of SN neutrino events for $0\leq t_{pb}\leq 100$ ms for $\nu_e{\rm Ar}$ by DUNE detector and IBD signals by JUNO and HyperK detectors in different flavor transition scenarios for simulations G, B, F, and N with progenitor masses of $8.8$, $10$, $11.2$, and $13~{\rm M}_\odot$, respectively. }
	\label{totnumtpb}
\end{table}

Table~\ref{totnumtpb} presents the total event numbers obtained from the event rates shown in Fig.~\ref{duneAr} as well as total event numbers of IBD detection channel of SN neutrinos as will be discussed momentarily. 
The ordering of total $\nu_e{\rm Ar}$
event numbers follows that of the peak event rate, i.e., VFT$>$FE$>$MSW-IO$>$MSW-NO for simulations G, B, and F while MSW-NO$>$ MSW-IO$>$FE$>$VFT for simulation N.

Besides $\nu_e{\rm Ar}$ signals, SN neutrinos also interact in the following channels in liquid argon time projection chamber detectors: $\nu + e^- \rightarrow\nu + e^-$ and $\bar{\nu}_e{\rm Ar}$ charged-current interaction, $\bar{\nu}_e+{^{40}{\rm Ar}} \rightarrow {^{40}{\rm Cl}^*+e^+}$. The event rates of these channels are subdominant compared to $\nu_e{\rm Ar}$ interactions (for a reference, see Table II in \cite{Lujan-Peschard:2014lta} and Table I in \cite{Lu:2016ipr}). Therefore, we only focus on $\nu_e{\rm Ar}$ interactions.

\begin{figure}[htbp]
	\begin{center}
		\includegraphics[width=.5\columnwidth]{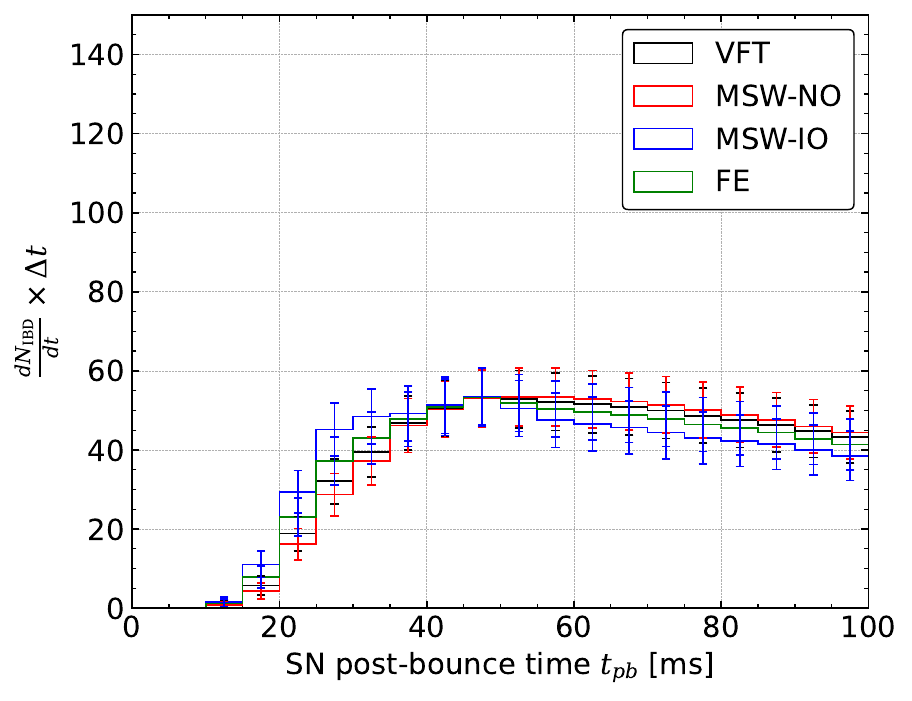}%
		\includegraphics[width=.5\columnwidth]{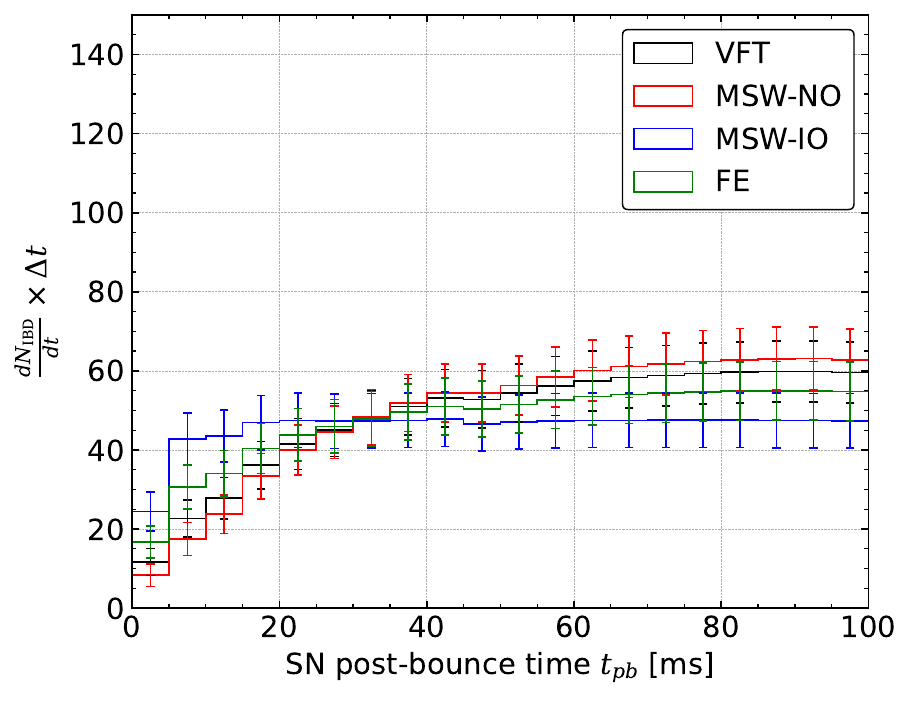}\\
		\includegraphics[width=.5\columnwidth]{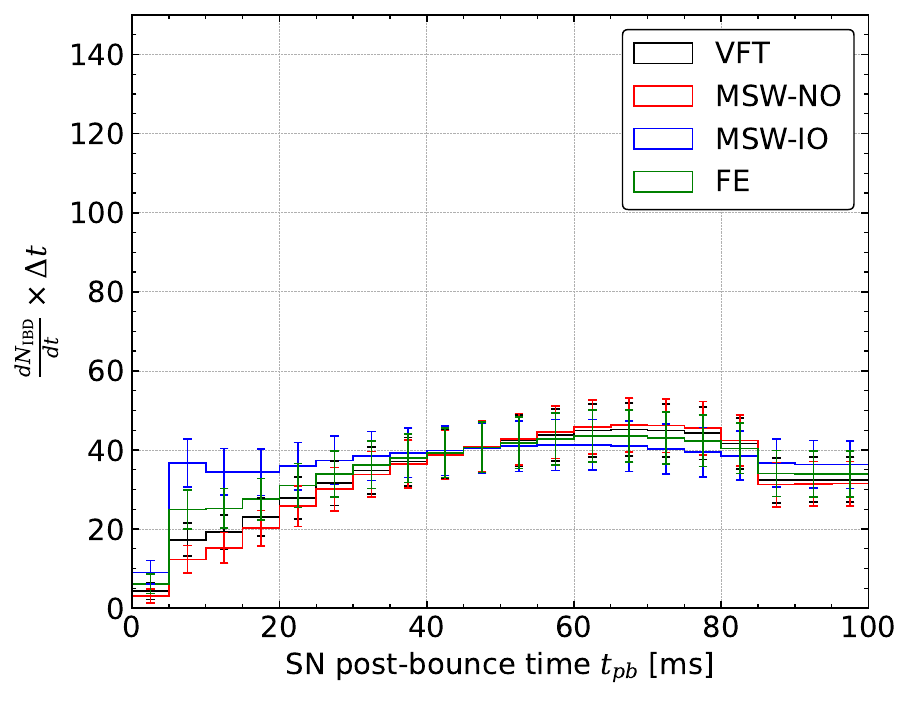}%
		\includegraphics[width=.5\columnwidth]{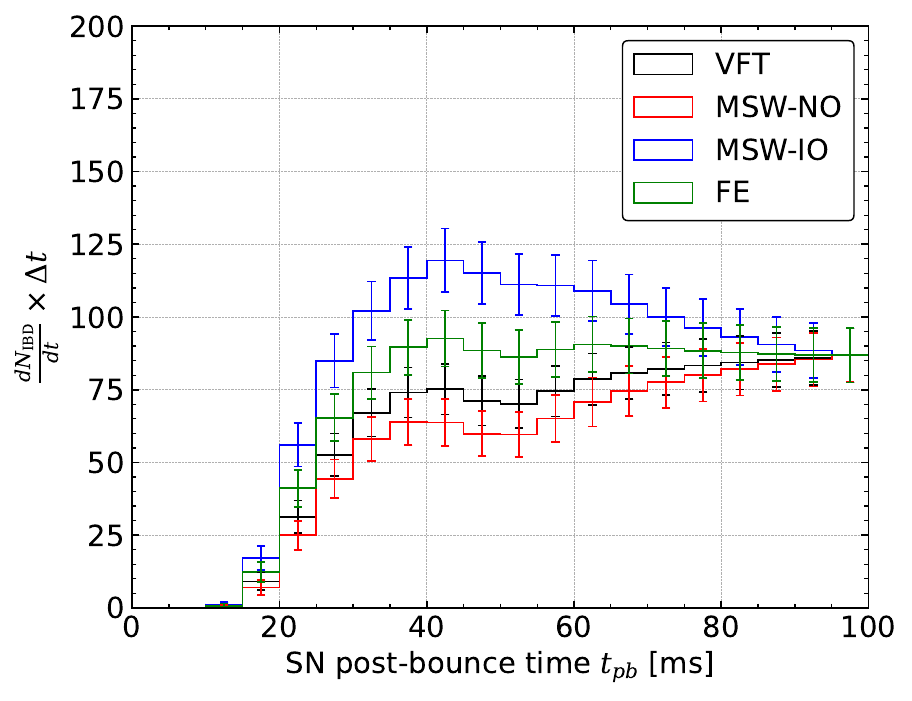}%
		\caption{IBD event rates in JUNO detector for different flavor transition scenarios. 
		These event rates are obtained for simulations G, B, F, and N (arranged in the same way as Fig.~\ref{luminosity}) with progenitor masses of $8.8$, $10$, $11.2$, and $13~{\rm M}_\odot$, respectively. }
		\label{junoIBD}
	\end{center}
\end{figure}

In JUNO scintillation detector, the spectrum of IBD events is obtained by measuring the positron energy deposit. The predicted event spectrum is given by
\begin{eqnarray}
	&& \left(\frac{d^2N_{\rm IBD}}{dE_{e^+}dt}\right)\Delta t=N_p\cdot\int dE_\nu F_{\bar{e}}(E_{\nu},t)\Delta t\cdot\frac{d\sigma_{\rm IBD}(E_\nu,~E_{e^+})}{dE_{e^+}}, \label{IBDspec} 
%	&& \left(\frac{dN_{{\rm IBD}}}{dt}\right)\Delta t=\left(\frac{dN_{e^+}}{dt}\right)\Delta t=N_p\cdot\int_{E_{\rm min}}^\infty dE_\nu\frac{dF_{\bar{e}}}{dE_\nu}\Delta t\cdot\sigma_{\rm IBD}(E_\nu),
\end{eqnarray}
where $N_p$ is the number of target protons in the detector, and $\sigma_{\rm IBD}(E_\nu)$ is the IBD cross section  taken from \cite{Strumia:2003zx}. The minimum neutrino energy for generating IBD interactions is $E_{\rm min}=1.8~{\rm MeV}$. Integrating the positron energy, we obtain the IBD event per $5$ ms, $(dN_{\rm IBD}/dt)\cdot \Delta t$, in JUNO \cite{An:2015jdp} as shown in Fig.~\ref{junoIBD}. 

IBD events are caused by $\bar{\nu}_e$ flux. In simulation N, $n_{\bar{\nu}_e}$ is comparable to $n_{\nu_x}$ %in the earlier time%
 as seen in Fig.~\ref{emsrate}. MSW oscillation does not change the situation since it involves only the full swapping (IO) or partial swapping (NO) between $\bar{\nu}_e$ and $\nu_x$. On the other hand, the mean energy of $\nu_x$ is significantly larger than that of $\bar{\nu}_e$ as described in Table~\ref{emission}. Since the IBD cross section grows with the neutrino energy, the full and partial swapping between $\bar{\nu}_e$ and $\nu_x$ in Eqs.~(\ref{ebarNH}) and (\ref{ebarIH}) imply that IBD event rate in MSW-IO scenario is larger than that of MSW-NO. 
In either VFT or FE scenario, the fraction of $\bar{\nu}_e$ resulting from the flavor transition of $\nu_x$ at the source is between MSW-IO and MSW-NO. Specifically, $F_{\bar{e}}$ contains $45\%$ and $67\%$ of $F_x^0$ for VFT and FE scenarios, respectively. Hence IBD event rates in simulation N follow the ordering MSW-IO$>$FE$>$VFT$>$MSW-NO.
For simulations G, B, and F, it is seen that $n_{\nu_x}>n_{\bar{\nu}_e}$ in the earlier time. Hence in the earlier time the IBD event rate in MSW-IO scenario is largest due to the complete swapping between $\nu_x$ and $\bar{\nu}_e$.

The IBD event spectrum in HyperK detector~\cite{Hyper-Kamiokande:2018ofw} can also be calculated by Eq.~(\ref{IBDspec}) with $N_p=2.48\times 10^{34}$, which is the number of free protons in the designed fiducial mass of $374$ ktons.  
To calculate the number of IBD events per $5$ ms, we integrate the positron energy with the energy resolution and the event threshold energy taken into account, i.e., 
 \begin{eqnarray}
 \left(\frac{dN_{\rm IBD}}{dt}\right)\Delta t=\int_{E_{\rm th}} dE_d \int dE_{e^+}\left(\frac{d^2N_{\rm IBD}}{dE_{e^+}dt}\right)\cdot R(E_d, E_{e^+})\Delta t,	
\label{HyperK}
\end{eqnarray}
where $R(E_d, E_{e^+})$ is the energy resolution function given by
 \begin{eqnarray}
R(E_d, E_{e^+})=\frac{1}{\Delta\sqrt{2\pi}}\exp\left(-\frac{(E_d-E_{e^+})^2}{2\Delta^2}\right),	
\label{energy_res}
\end{eqnarray} 
with $E_d$ the detected energy, $\Delta/{\rm MeV}\equiv 0.6\sqrt{E_d/{\rm MeV}}$ the energy resolution, and $E_{\rm th}\equiv (5 \ {\rm MeV}+m_e)$ the event threshold energy~\cite{Hyper-Kamiokande:2021frf}.  
The time profiles of IBD event rates in HyperK corresponding to different SN neutrino simulations and neutrino flavor transitions are similar to those in JUNO except the normalizations. We will not show these time profiles but point out that the HyperK IBD event number 
for $0\leq t_{pb}\leq 100$ ms for each SN neutrino emission simulation and flavor transition scenario is summarized in Table~\ref{totnumtpb}. It is approximately 17 times larger than that of JUNO regardless SN neutrino simulations and flavor transition scenarios.

\label{SNevents}

\section{Examining the Presence of MSW Effects}

\subsection{Cumulative time distribution}
To characterize the sharp rise of $\nu_e$ flux during the neutronization burst, we define cumulative time distributions of SN neutrino signals for the time interval of interest $t = (0-0.1) \ {\rm s}$ as in~\cite{Serpico:2011ir}
\begin{equation}
K^{i,{\rm Ar}}(t)=\frac{\int^t_0 \frac{dN^i_{\rm Ar}}{dt'}dt'}{\int^{0.1{\rm s}}_0 \frac{dN^i_{\rm Ar}}{dt'}dt'},
\label{cum_Ar}
\end{equation}
where $i={\rm VFT}$, MSW-NO, MSW-IO, and FE. 

We note that the authors in~\cite{Serpico:2011ir} introduced cumulative time distribution of SN $\bar{\nu}_e$ events and applied it to decipher the neutrino mass orderings with IceCube detector. In our study, we construct not only $K^{i,{\rm Ar}}(t)$ for liquid argon detectors but also $K^{i,{\rm IBD}}(t)$ for liquid scintillation detectors. Furthermore, we shall calculate the area under each of the above cumulative time distributions, which is a powerful diagnostic quantity in our study (see discussions later).       

To make comparisons between various time dependencies of SN event rates, it is important to
adopt an observational definition for the origin of time. For example, although the characteristic event peak (driven by $\nu_e$ flux) in VFT, FE and MSW-IO event rates most likely appears in $t_{\text{pb}}=(5-10)$ ms, the postbounce time is not an experimentally observable quantity. The time variable $t^*$ we propose is defined as follows. First we propose to bin the neutrino events with a $5$ ms bin width.
Second, if the characteristic sharp peak emerges (such as in MSW-IO, FE and VFT scenarios), 
then $t^*=0$ is defined at $5$ ms before the peak. In other words the end of the first time bin, $t^*=5$ ms, is exactly at the beginning of the peak. Finally, if the sharp peak does not appear, then $t^*=0$ is defined at the time when first event appears. For the IBD events to be discussed later, we also define the origin of $t^*$ in this way since there are no clear event peaks as shown by Fig.~\ref{junoIBD}. It is important to note that the origin of $t^*$ depends on simulations. Taking Fig.~\ref{junoIBD} as an example, the SN neutrino events predicted by simulations G and N occur significantly later than those predicted by simulations B and F.  

\begin{figure}[H]%
	\begin{center}%
	\includegraphics[width=.5\columnwidth]{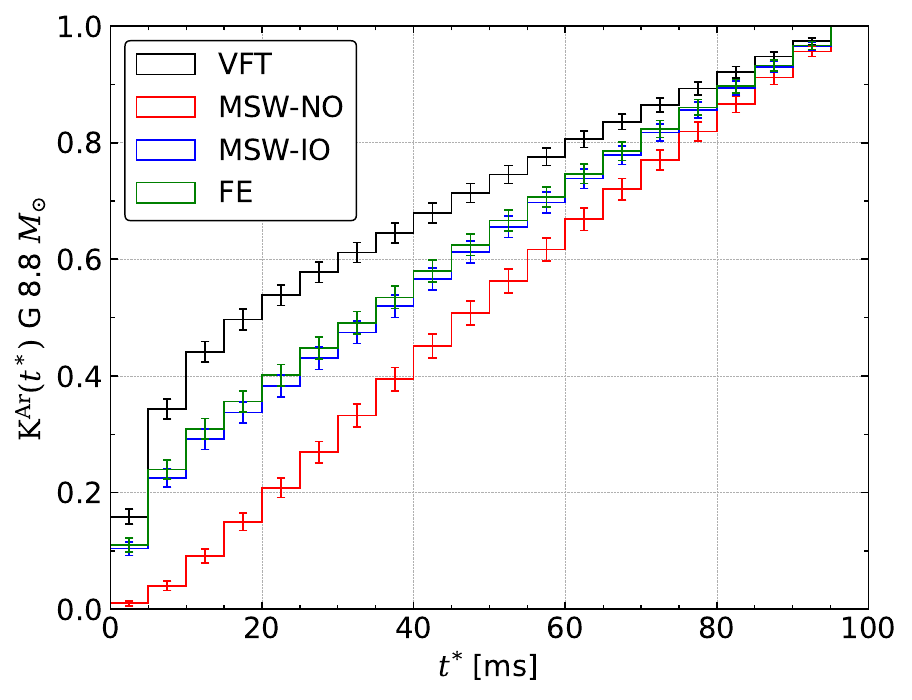}%
	\includegraphics[width=.5\columnwidth]{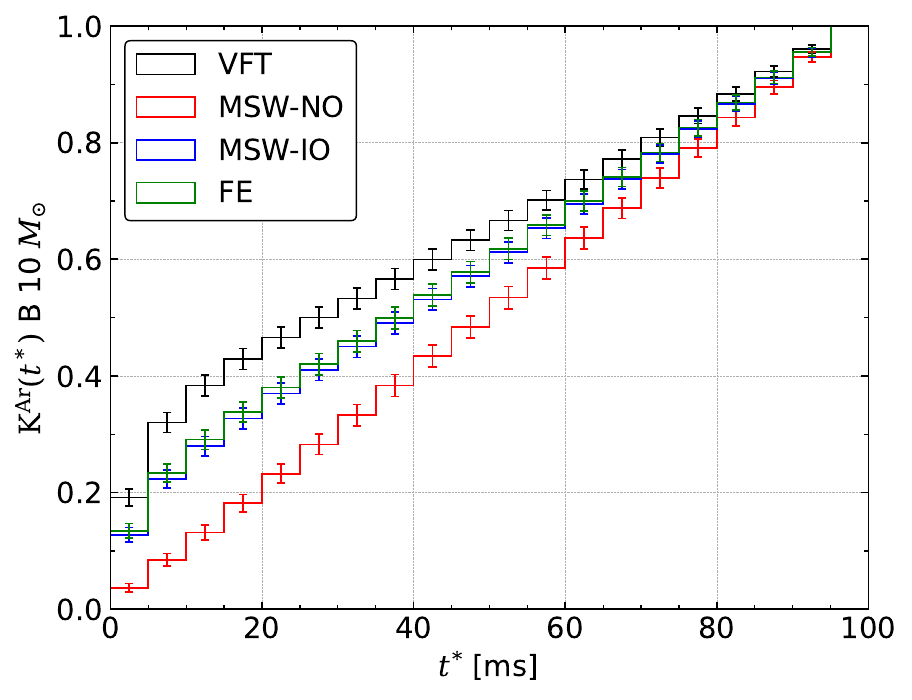}\\
	\includegraphics[width=.5\columnwidth]{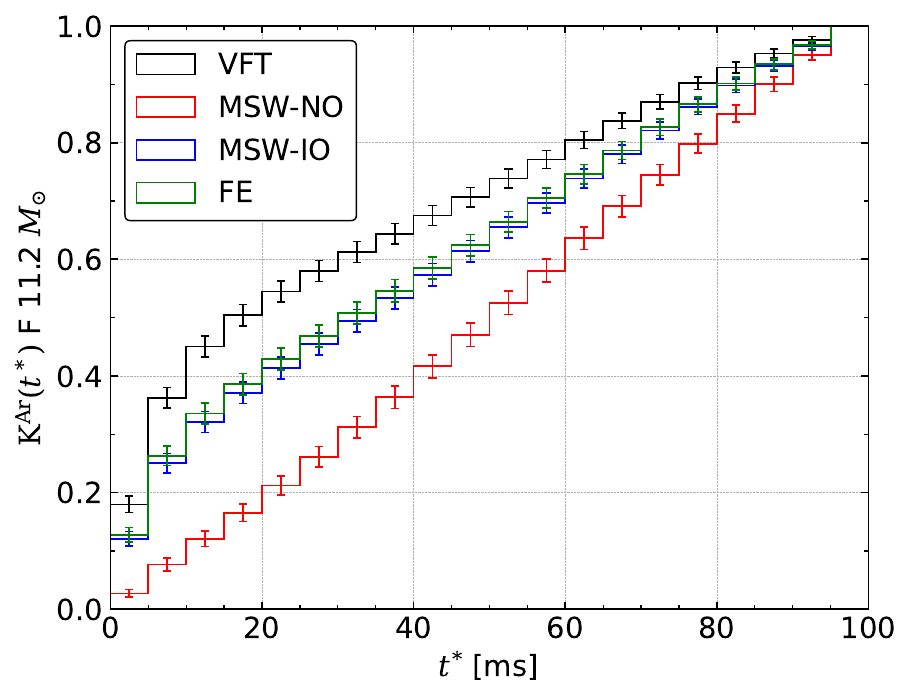}%
	\includegraphics[width=.5\columnwidth]{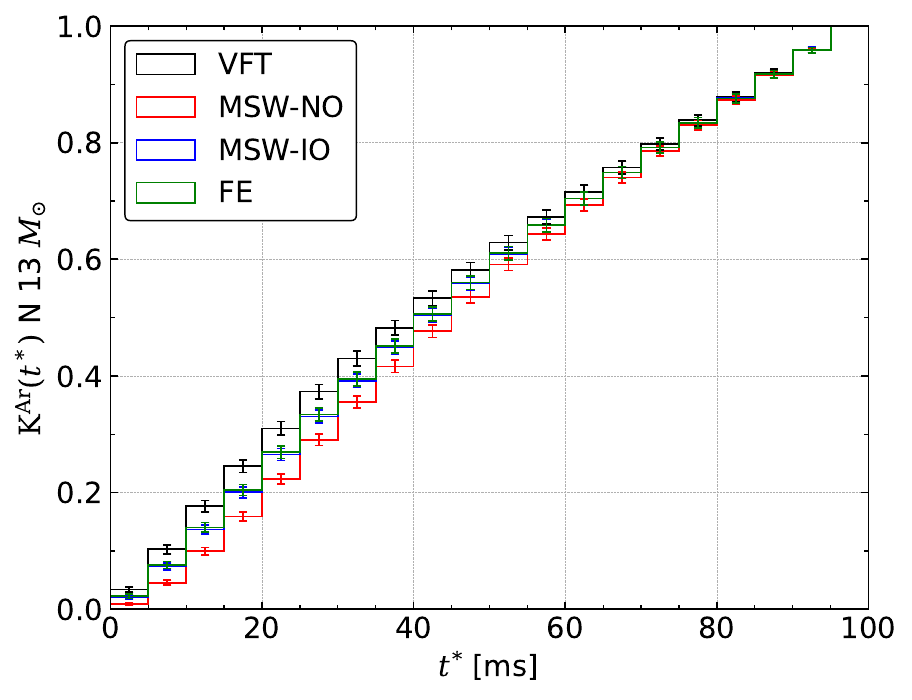}%
	\caption{Cumulative time distributions $K^{i,{\rm Ar}}(t^*)$ of $\nu_e{\rm Ar}$ signals in DUNE detector in the time period of $0\leq t^* \leq 0.1~{\rm s}$. These event distributions are predictions of simulations G, B, F, and N for progenitor masses of $8.8$, $10$, $11.2$, and $13~{\rm M}_\odot$, respectively.}%
	\label{cumul_Ar}%
	\end{center}%
\end{figure}

In Fig.~\ref{cumul_Ar}, we present $K^{i,{\rm Ar}}(t^*)$, the cumulative time distributions of $\nu_e{\rm Ar}$ event rates, in VFT, MSW-NO, MSW-IO, and FE scenarios, respectively. By definition, $0\leq K^{i,{\rm Ar}}(t^*)\leq 1$. The method for calculating the statistical errors of $K^{i,{\rm Ar}}(t^*)$ as well as $K^{i,{\rm IBD}}(t^*)$ to be defined momentarily is discussed in Appendix A. In simulations G, B, and F, the neutronization peak causes $K^{\rm{VFT,Ar}}(t^*)$, $K^{\rm{FE,Ar}}(t^*)$, and $K^{\rm{IO,Ar}}(t^*)$ to increase faster in the beginning while $K^{\rm{NO,Ar}}(t^*)$ increases in a more even pace. Quantitatively, the ordering $K^{\rm VFT, Ar}(t^*)>K^{\rm FE, Ar}(t^*)>K^{\rm IO, Ar}(t^*)>K^{\rm NO, Ar}(t^*)$ holds in these simulations. 
%The first inequality arises because VFT scenario produces the highest $\nu_e{\rm Ar}$ event peak. 
Clearly $K^{i,{\rm Ar}}(t^*)$ is useful for discriminating different flavor transition scenarios due to the significant differences between peak event rates in different scenarios. 
In simulation N, $K^{i,{\rm Ar}}(t^*)$ is much less useful for discriminating different scenarios since none of the flavor transition scenarios exhibit sharp neutronization peaks.  

\begin{figure}[H]
	\begin{center}
	\includegraphics[width=.5\columnwidth]{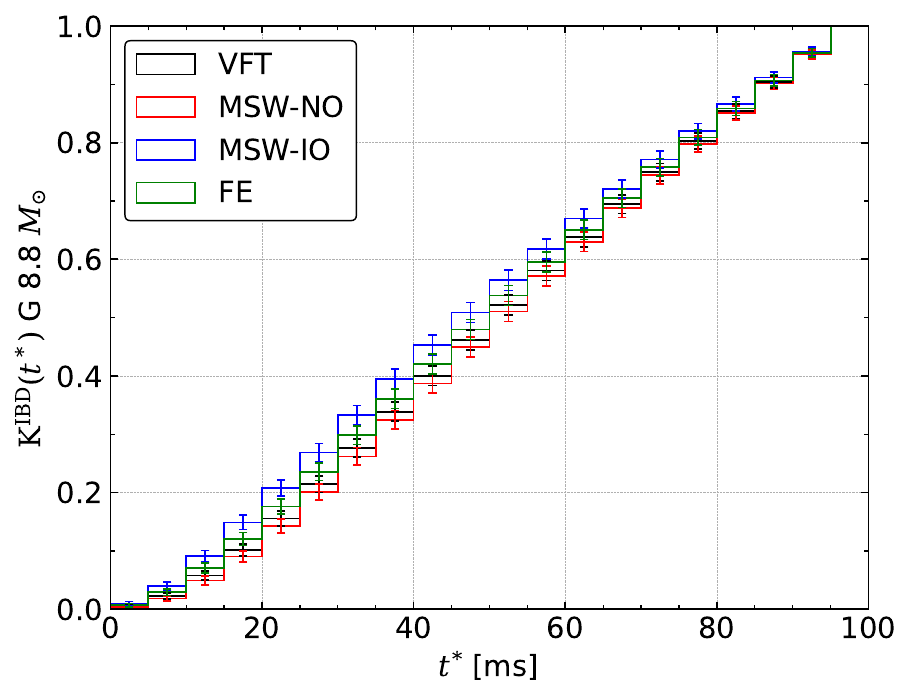}%
	\includegraphics[width=.5\columnwidth]{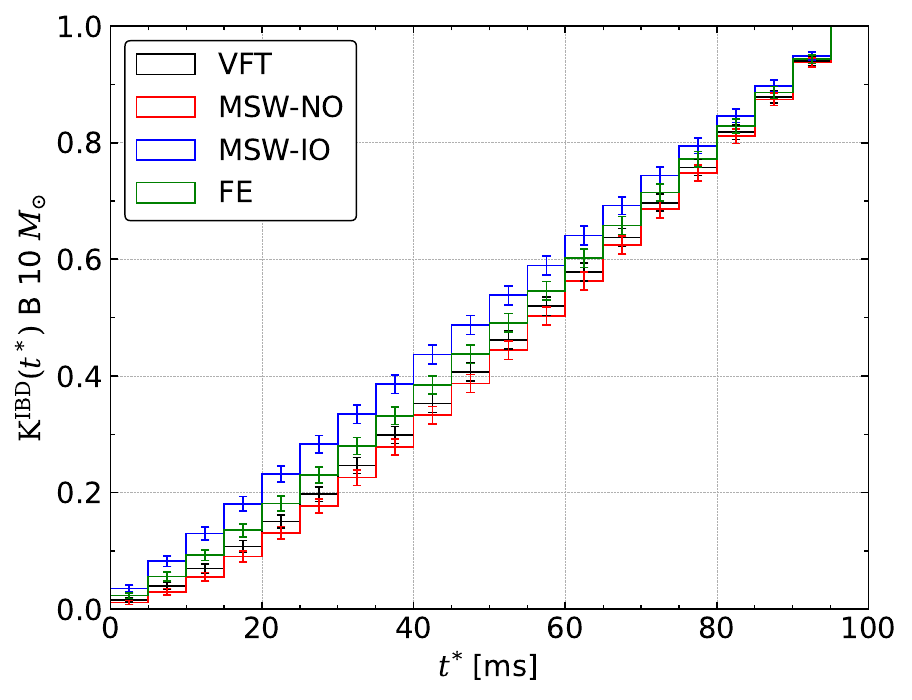}\\
	\includegraphics[width=.5\columnwidth]{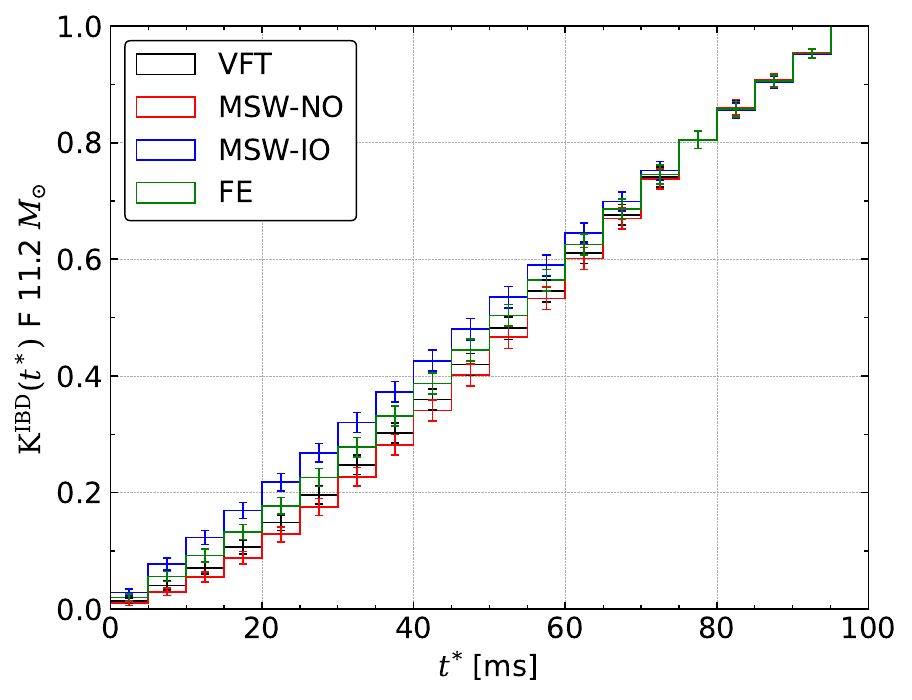}%
	\includegraphics[width=.5\columnwidth]{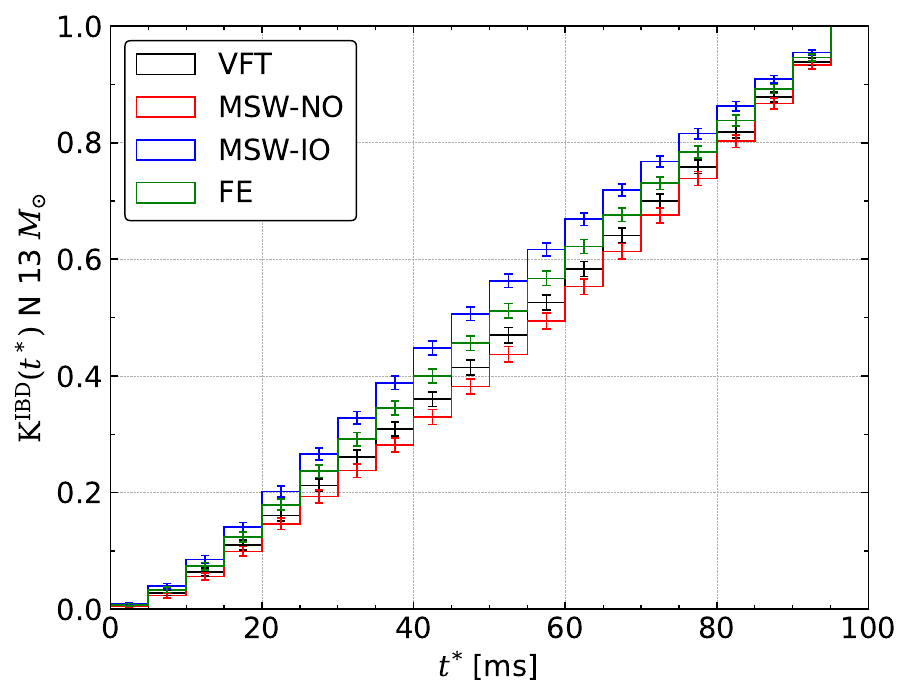}%
	\caption{Cumulative time distributions $K^{i,{\rm IBD}}(t^*)$ of $\bar{\nu}_e$ IBD events in JUNO detector in the time period of $0\leq t^* \leq 0.1~{\rm s}$. These event distributions are predictions of simulations G, B, F, and N for progenitor masses of $8.8$, $10$, $11.2$, and $13~{\rm M}_\odot$, respectively.}
	\label{cumul_IBD}
	\end{center}
\end{figure}

Similar cumulative time distributions can be defined for IBD events. We present $K^{i,{\rm IBD}}(t^*)$ in Fig.~\ref{cumul_IBD}. For all simulations considered, it is clear that $K^{\rm IO,IBD}(t^*)$ is largest among all scenarios since IBD event rate of MSW-IO increases the fastest in the earlier time. Explicitly, the ordering $K^{\rm IO,IBD}(t^*)>K^{\rm FE,IBD}>K^{\rm VFT,IBD}(t^*)>K^{\rm NO,IBD}(t^*)$ holds although differences among $K^{\rm FE,IBD}$, $K^{\rm VFT,IBD}$ and $K^{\rm NO,IBD}$ are not significant. 

In the next subsection, we shall quantify the time profiles of $K^{i,{\rm Ar}}(t^*)$ and  $K^{i,{\rm IBD}}(t^*)$ for different flavor transition scenarios.
%We shall see momentarily that VFT and MSW-NO can be distinguished by simply invoking $\nu_e$Ar events. On the the hand, to distinguish between VFT and MSW-IO, both $\nu_e$Ar and $\bar{\nu}_e$ IBD events are required. 

\subsection{Integral of cumulative time distribution }

\begin{figure}[H]
    \centering
    \includegraphics[width=.5\columnwidth]{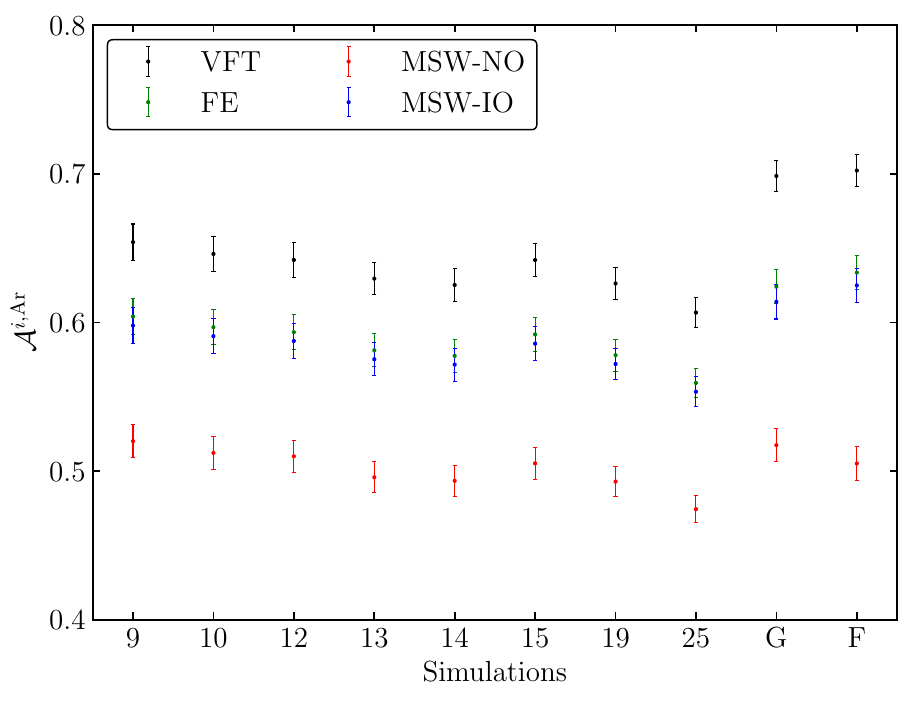}%
    \includegraphics[width=.5\columnwidth]{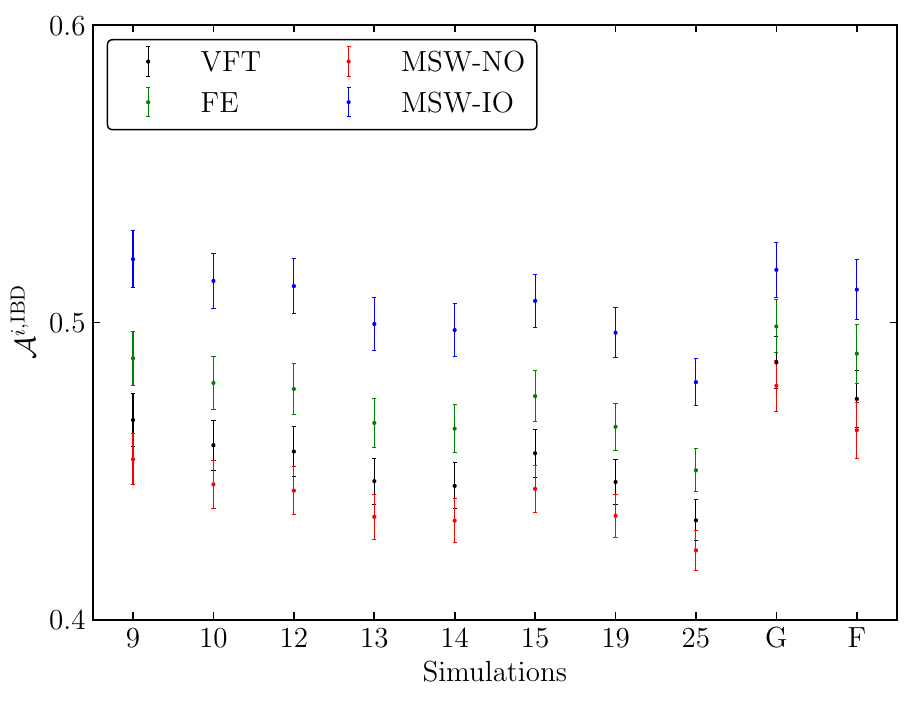}\\%
    \includegraphics[width=.5\columnwidth]{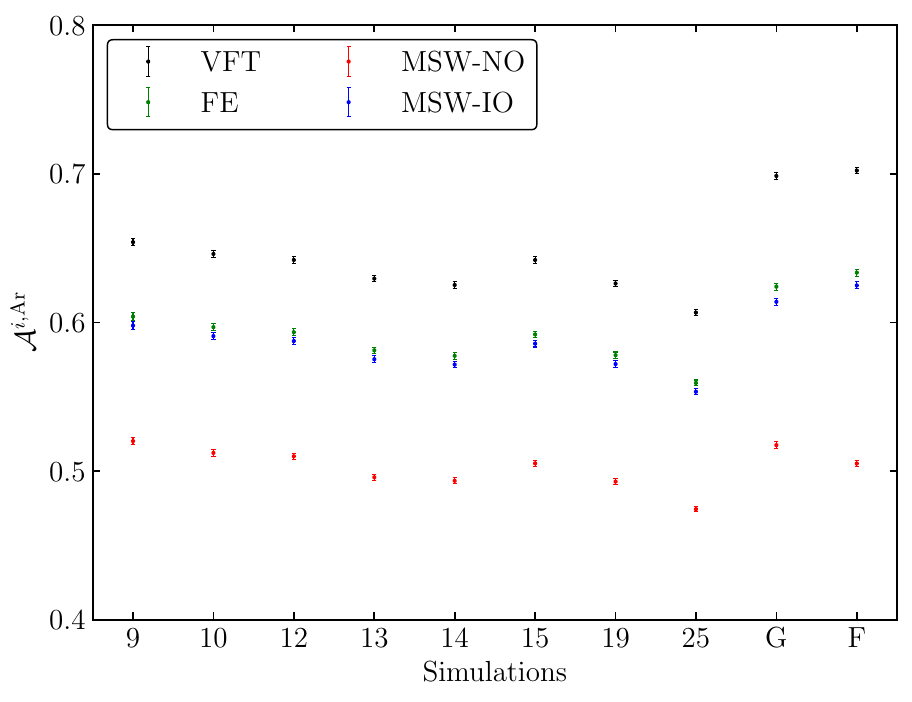}%
    \includegraphics[width=.5\columnwidth]{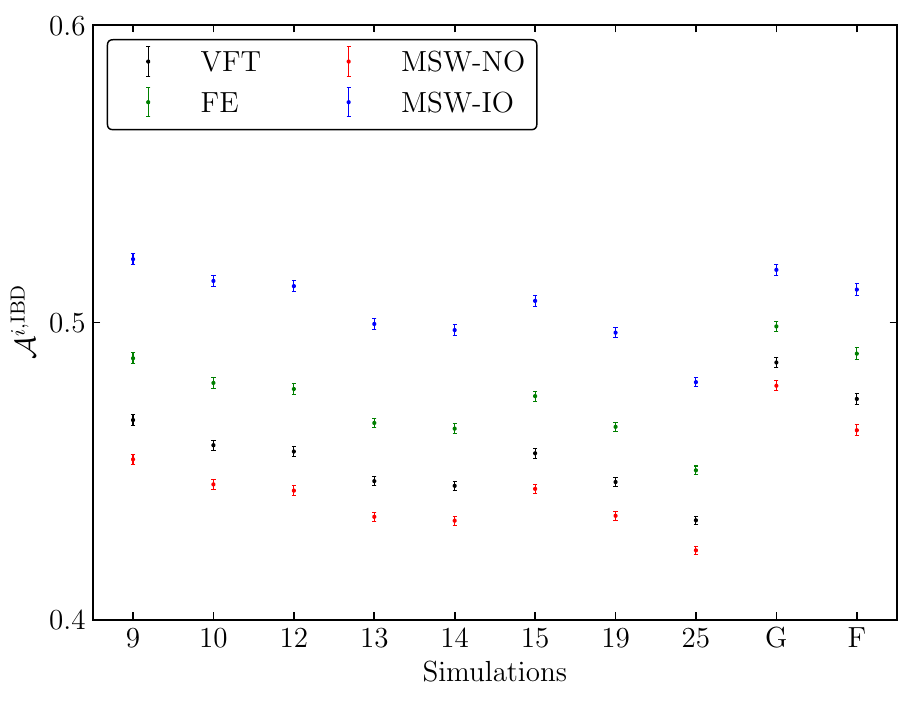}\\%
    \includegraphics[width=.5\columnwidth]{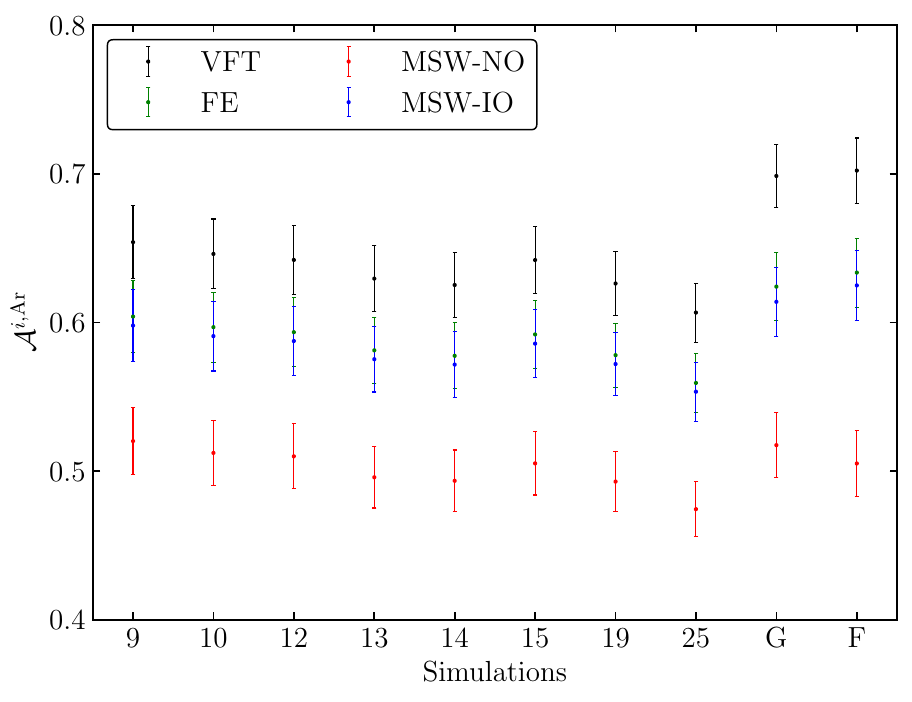}%
    \includegraphics[width=.5\columnwidth]{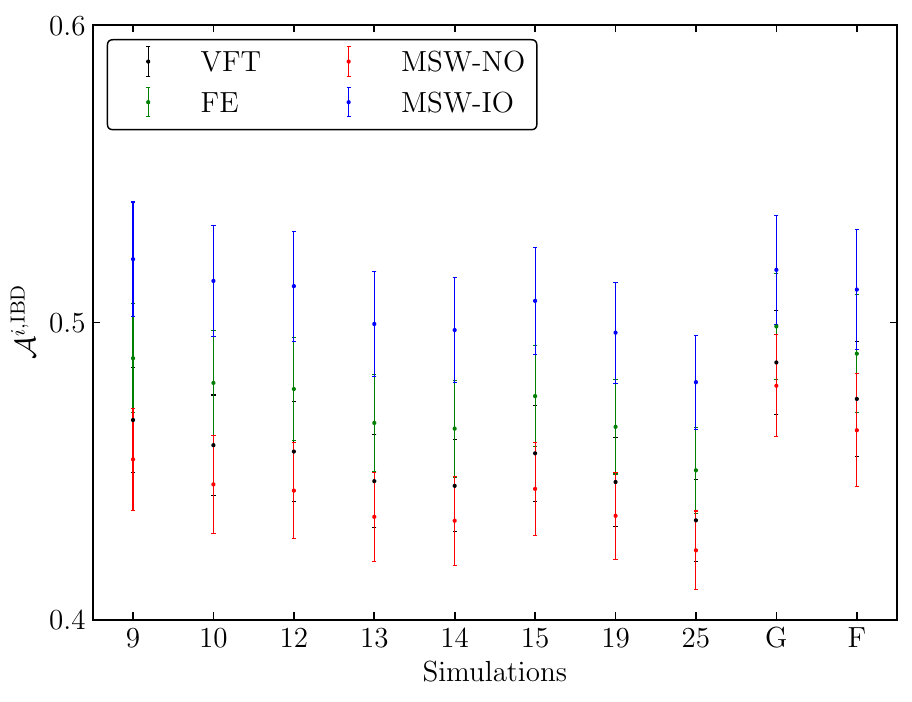}%
    \caption{Integrals of cumulative time distributions $\mathcal{A}^{i,{\rm Ar}}$ of $\nu_e{\rm Ar}$ events in DUNE detector (left panels) and $\mathcal{A}^{i,{\rm IBD}}$ of IBD events in JUNO detector (right panels). Top panel shows results for the benchmark SN distance of $d=5 \ {\rm kpc}$, while middle and bottom panels show results 
for  $d=1 \ {\rm kpc}$ and $d=10 \ {\rm kpc}$, respectively.   
    G and F on the $x$ axis denote predictions by simulations G and F corresponding to progenitor masses of $8.8$ and $11.2~{\rm M}_\odot$, respectively, while the numbers denote predictions by  simulation B for different progenitor masses in the unit of ${\rm M}_{\odot}$.}
    \label{inte_cumu}
\end{figure}

For experimental analysis in liquid argon detector, we quantify the ordering of $K^{i,{\rm Ar}}(t^*)$ by integrating each $K$ over the time period of interests. Explicitly, the integral of cumulative time distribution ($\mathcal{A}$) is given by
\begin{equation}
    \mathcal{A}^{i,{\rm Ar}}=\frac{1}{T}\int^T_0 K^{i,{\rm Ar}}(t^*)dt^*,
    \label{Icd_Ar}
\end{equation}
with $T=0.1{\rm s}$. Similar integral can be defined for cumulative time distribution of IBD events. We note that $\mathcal{A}^{i,{\rm Ar (IBD)}}$ is the normalized area under the cumulative time distribution $K^{i,{\rm Ar  (IBD)}}(t^*)$. The ordering of $K^{i,{\rm Ar  (IBD)}}(t^*)$ for different flavor transition scenarios is naturally preserved in the ordering of $\mathcal{A}^{i,{\rm Ar (IBD)}}$. Moreover, the latter is a more convenient diagnostic quantity since it is just a number. In Fig.~\ref{inte_cumu}, we present values of $\mathcal{A}$ for DUNE and JUNO detectors in different flavor transition scenarios predicted by simulations G, F, and B for SN distances 
$d=5$, $1$, and $10 \ {\rm kpc}$, respectively. For simulation B, we include results from different progenitor masses labeled as numbers in the unit of ${\rm M}_{\odot}$ as shown on the $x$ axis of each plot. The statistical error of $\mathcal{A}$ is calculated using the formula in the Appendix. 
We do not consider simulation N from this stage on because it gives very different predictions on neutrino flux spectra, particularly on the neutronization peak of $\nu_e$ flux. In simulation N, this peak is much broader and lower than those predicted by other simulations. 
We  do not show $\mathcal{A}^{i,{\rm IBD}}$ for HyperK since they differ from those of JUNO only in statistical uncertainties. 
However these results will be included later when we combine  $\mathcal{A}^{i,{\rm IBD}}$ with $\mathcal{A}^{i,{\rm Ar}}$ for testing MSW flavor transitions of SN neutrinos. 
%As we shall see later,  the reducing errors are not helpful for distinguishing different flavor transition scenarios 
%since $\mathcal{A}^{i,{\rm IBD}}$ varies significant with the simulation.  
%their central values are very similar to those in JUNO. Although $\mathcal{A}^{i,{\rm IBD}}$ in HyperK are smaller due to a larger statistics ($\sim 17$ times more IBD events than JUNO),  
 
On left panels of Fig.~\ref{inte_cumu}, one can see that $\mathcal{A}^{\rm VFT,Ar}>\mathcal{A}^{\rm FE,Ar}> \mathcal{A}^{\rm IO,Ar}>\mathcal{A}^{\rm NO,Ar}$ for simulations G, B, and F. For central values, $\mathcal{A}^{\rm NO,Ar}\leq 0.52$ while $\mathcal{A}^{\rm IO,Ar}\geq 0.55$, and $\mathcal{A}^{\rm VFT,Ar}\geq 0.60$. The central value of 
$\mathcal{A}^{\rm FE,Ar}$ is slightly larger than that of $\mathcal{A}^{\rm IO,Ar}$.
Looking at the plot for $\mathcal{A}^{i,{\rm Ar}}$ for $d=5 \ {\rm kpc}$, it is seen that MSW-NO can in general be distinguished from other scenarios, except that $\mathcal{A}^{\rm NO,Ar}$ of simulation G ($8.8 \ {\rm M}_{\odot}$ progenitor mass) and $\mathcal{A}^{\rm IO,Ar}$ of simulation B with $25 \ {\rm M}_{\odot}$ progenitor mass almost overlap. On the other hand, with similar progenitor masses, $\mathcal{A}^{\rm IO,Ar}$ of simulation B with $9.0 \ {\rm M}_{\odot}$ progenitor mass is well separated from $\mathcal{A}^{\rm NO,Ar}$ of simulation G. We also observe that the separation between MSW-NO and VFT is quite significant. Finally, one can see that $\mathcal{A}^{\rm IO,Ar}$ predicted by simulations G and F overlaps with $\mathcal{A}^{\rm VFT,Ar}$ predicted by simulations B for several progenitor masses. Therefore, VFT and MSW-IO are not distinguishable by $\nu_e {\rm Ar}$ events alone. Since $\mathcal{A}^{\rm FE,Ar}$ is in between $\mathcal{A}^{\rm VFT,Ar}$ and  $\mathcal{A}^{\rm IO,Ar}$, one also concludes that FE and MSW-IO are not distinguishable by relying on $\nu_e {\rm Ar}$ events only.
For $d=1 \ {\rm kpc}$, the separation between MSW-NO and other scenarios become very apparent. For $d=10 \ {\rm kpc}$, $\mathcal{A}^{\rm NO,Ar}$ of simulation G slightly overlaps with $\mathcal{A}^{\rm IO,Ar}$ of simulation B with $25 \ {\rm M}_{\odot}$ progenitor mass. 

On right panels, it is seen that  $\mathcal{A}^{\rm IO,IBD}>\mathcal{A}^{\rm FE,IBD}>\mathcal{A}^{\rm VFT,IBD}>\mathcal{A}^{\rm NO,IBD}$. For $d=5 \ {\rm kpc}$, $\mathcal{A}^{\rm VFT,IBD}$ predicted by simulation G overlaps with $\mathcal{A}^{\rm IO,IBD}$ predicted by simulation B for a few different progenitor masses, i.e., one cannot distinguish between MSW-IO and VFT with IBD events alone. 
Since $\mathcal{A}^{\rm FE,IBD}$ is in between $\mathcal{A}^{\rm IO,IBD}$ and $\mathcal{A}^{\rm VFT,IBD}$, MSW-IO and FE are also not distinguishable by relying on IBD events only. For $d=1 \ {\rm kpc}$, the separabilities between different flavor transition scenarios do not improve.
In order to discriminate between MSW-IO and VFT, we observe that $\mathcal{A}^{\rm VFT,Ar}>\mathcal{A}^{\rm IO,Ar}$ while $\mathcal{A}^{\rm IO,IBD}>\mathcal{A}^{\rm VFT,IBD}$ for all simulations.
Similarly, for all simulations, we also observe that $\mathcal{A}^{\rm FE,Ar}>\mathcal{A}^{\rm IO,Ar}$ while  $\mathcal{A}^{\rm IO,IBD}>\mathcal{A}^{\rm FE,IBD}$.  This motivates us to use the ratio
\begin{equation}
    \mathcal{R}^i\equiv\frac{\mathcal{A}^{i,{\rm Ar}}}{\mathcal{A}^{i,{\rm IBD}}}
    \label{ratio_Icd}
\end{equation}
for discriminating between MSW-IO and VFT or between MSW-IO and FE. 
%For $d=1 \  {\rm kpc}$ and  $d=10 \  {\rm kpc}$, one can neither discriminate between MSW-IO and VFT  nor discriminate between MSW-IO and FE by IBD events alone.
%It remains necessary to invoke $\mathcal{R}^i$ for making the above discriminations. 

%Before moving on, we note that the statistical error of each $\mathcal{A}^{i,{\rm IBD}}$ in HyperK is much smaller than that of JUNO since the former can detect approximately $17$ times more IBD events.  
%By considering IBD events only, the reduced error is not helpful for distinguishing different flavor transition scenarios since $\mathcal{A}^{i,{\rm IBD}}$ varies significantly with the simulation.       

\begin{figure}[H]
    \centering
    \includegraphics[width=.5\columnwidth]{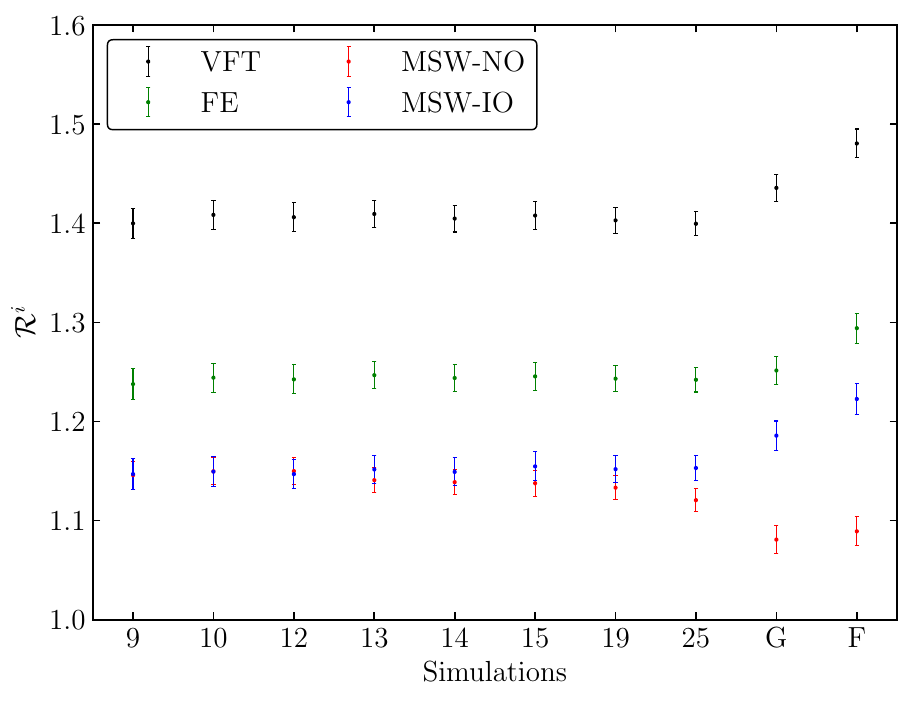}%
    \includegraphics[width=.5\columnwidth]{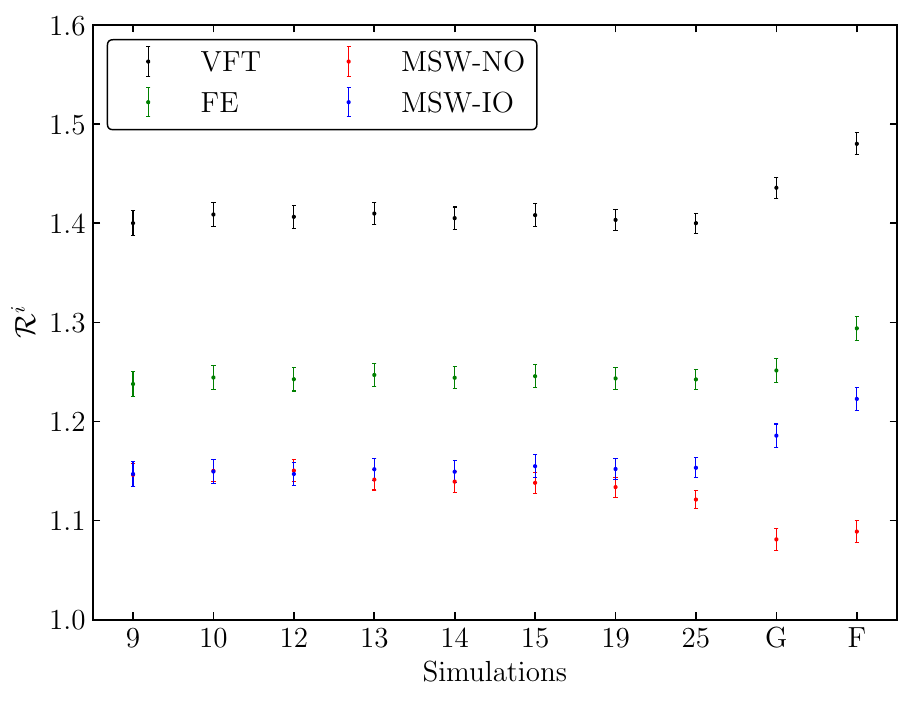}\\%
    \includegraphics[width=.5\columnwidth]{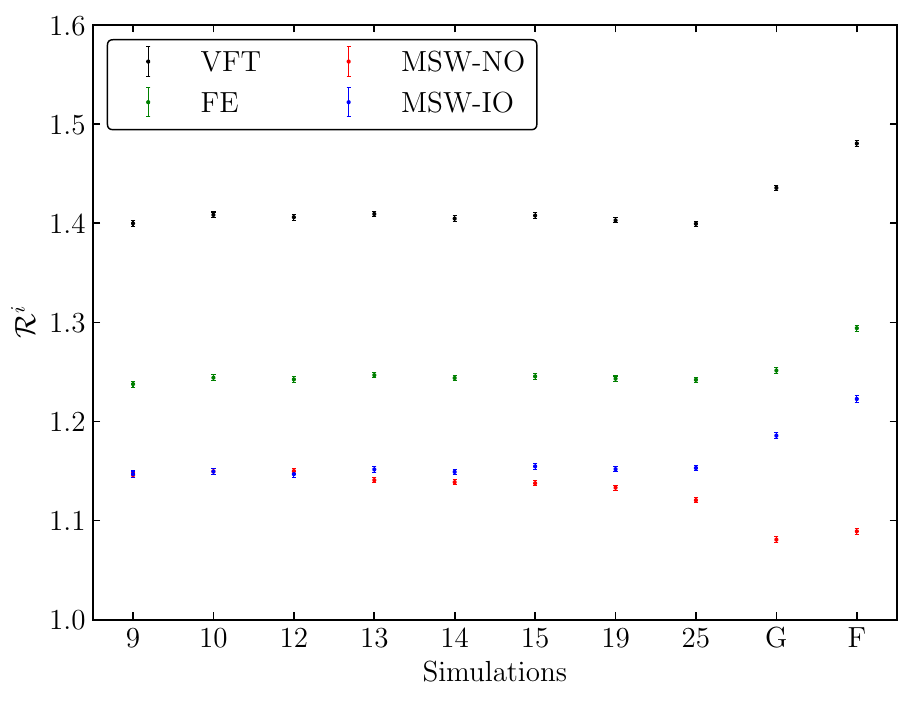}%
    \includegraphics[width=.5\columnwidth]{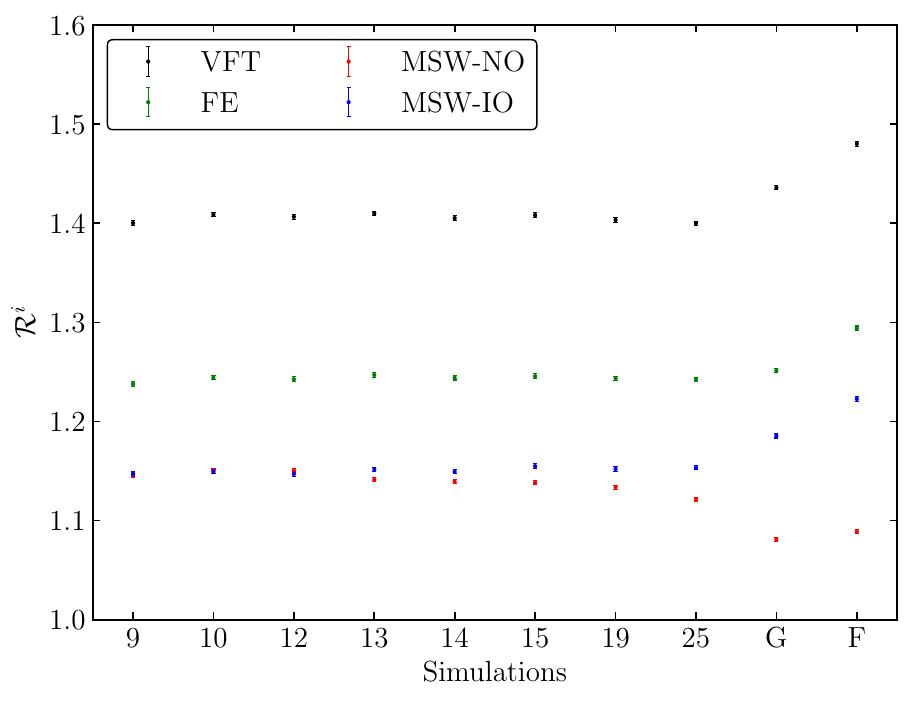}\\%
    \includegraphics[width=.5\columnwidth]{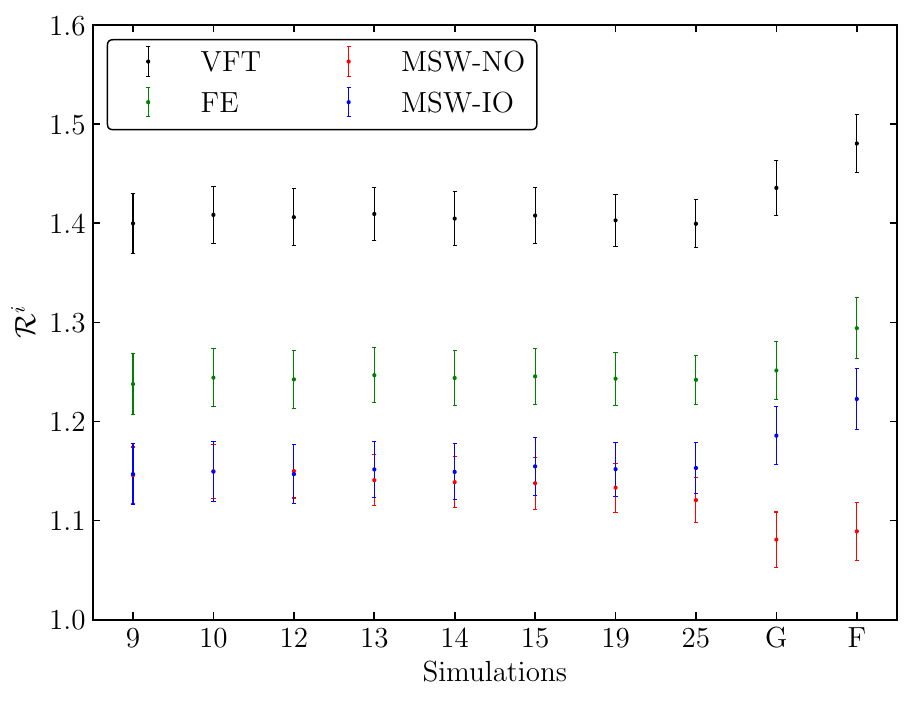}%
    \includegraphics[width=.5\columnwidth]{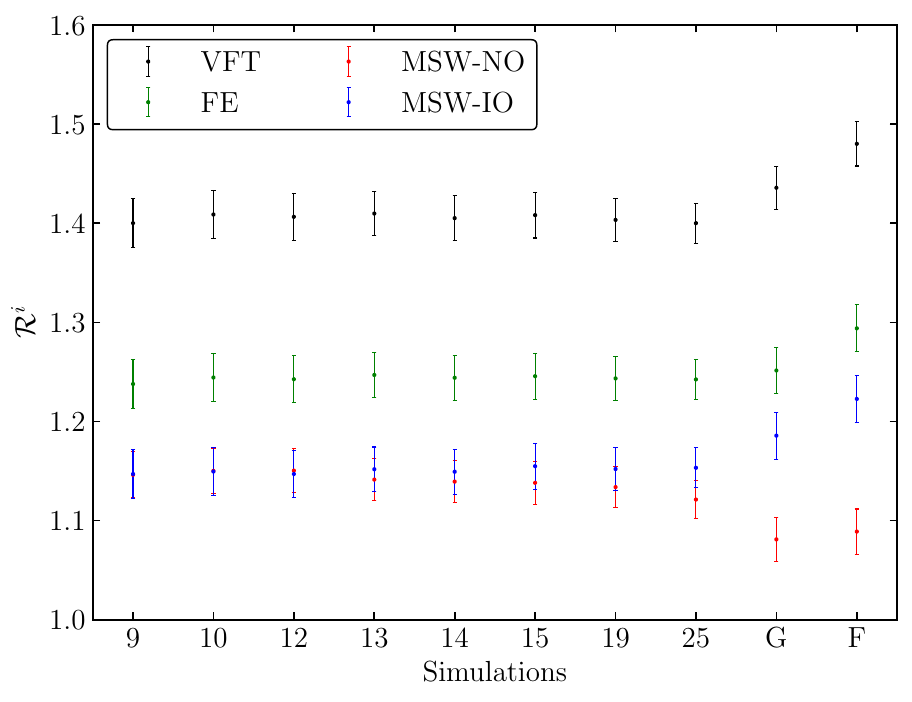}%
    \caption{%Ratios of the Integrated Cumulative distributions 
 Values of $\mathcal{R}^i$ with the denominator $\mathcal{A}^{i,{\rm IBD}}$ given by JUNO (left panels) and HyperK (right panels), respectively. From top to bottom, we take the SN distance as  
$5$, $1$, and $10 \ {\rm kpc}$, respectively.}
    \label{rIcd}
\end{figure}

In Fig.~\ref{rIcd}, we present values of $\mathcal{R}^i$ with the denominator $\mathcal{A}^{i,{\rm IBD}}$ given by JUNO (left panels) and HyperK (right panels), respectively. From top to bottom, we take the SN distance as  
$5$, $1$, and $10 \ {\rm kpc}$, respectively. In Fig.~\ref{rIcd}, it is seen that the red bars representing VFT scenarios are well separated from the blue ones representing MSW-IO
on both left and right panels, and for all three SN distances. Quantitatively speaking, for central values $\mathcal{R}^{\rm VFT}\geq1.40$ while $\mathcal{R}^{\rm IO}\leq1.22$ for all simulations. The statistical uncertainties are too small to account for the differences between $\mathcal{R}^{\rm VFT}$ and $\mathcal{R}^{\rm IO}$. Hence $\mathcal{R}^{i}$ is an effective quantity for discriminating VFT from MSW-IO.
We also observe that the values of  $\mathcal{R}^{\rm FE}$ are in between $\mathcal{R}^{\rm VFT}$ and $\mathcal{R}^{\rm IO}$. For $d=1 {\rm kpc}$, FE can also be separated from MSW-IO due to small statistical uncertainties. Finally it is interesting to see that for all three SN distances, VFT and FE are clearly separable no matter the denominator $\mathcal{A}^{i,{\rm IBD}}$ is from JUNO or HyperK.
\subsection{Earth matter effects}

Here we investigate whether Earth matter effects could affect discriminations between MSW and other flavor transition scenarios or not. As mentioned earlier, neutrinos arriving on Earth are in mass eigenstates~\cite{Dighe:1999bi}. Earth matter effects then modify the probability of each mass eigenstate being measured as a specific flavor eigenstate. Hence we have 
\begin{eqnarray}
P(\nu_1\to \nu_e)&= &|U_{e1}|^2-f_{\rm reg}, \nonumber \\
P(\nu_2\to \nu_e)&= &|U_{e2}|^2+f_{\rm reg},
\label{nu_regeneration}
\end{eqnarray}
%where we have neglected ${\cal O}(\sin^2\theta_{13})$  contributions in the above equations. 
The quantity $f_{\rm reg}$ is the regeneration factor resulting from Earth matter effects and its value depends on both the neutrino energy and the neutrino path length inside the Earth~\cite{deHolanda:2004fd}. For antineutrinos, we have  
\begin{eqnarray}
P(\bar{\nu}_1\to \bar{\nu}_e)&= &|U_{e1}|^2-\bar{f}_{\rm reg}, \nonumber \\
P(\bar{\nu}_2\to \bar{\nu}_e)&= &|U_{e2}|^2+\bar{f}_{\rm reg},
\label{anti-nu_regeneration}
\end{eqnarray}
where $\bar{f}_{\rm reg}$ is the regeneration factor with respect to antineutrinos. The analytic forms for $f_{\rm reg}$ and $\bar{f}_{\rm reg}$ have been given in~\cite{deHolanda:2004fd}, while Earth matter effects to the observations of SN neutrinos with different detectors have been discussed in~\cite{Dighe:2003jg,Dighe:2003vm,Borriello:2012zc,Liao:2016uis}.  
With Earth matter effects, the flavor transition probability of VFT scenario, Eq.~(\ref{incoherent}), is modified into
\begin{eqnarray}
P^m(\nu_{\beta} \to \nu_{\alpha})&=&(|U_{\alpha 1}|^2-f_{\rm reg}) |U_{\beta 1}|^2+(|U_{\alpha 2}|^2+f_{\rm reg}) |U_{\beta 2}|^2+|U_{\alpha 3}|^2 |U_{\beta 3}|^2 \nonumber \\
&=&P(\nu_{\beta} \to \nu_{\alpha})+f_{\rm reg}(|U_{\beta 2}|^2-|U_{\beta 1}|^2)
\end{eqnarray}
for neutrinos, and
\begin{eqnarray}
P^m(\bar{\nu}_{\beta} \to \bar{\nu}_{\alpha})&=&(|U_{\alpha 1}|^2-\bar{f}_{\rm reg}) |U_{\beta 1}|^2+(|U_{\alpha 2}|^2+\bar{f}_{\rm reg}) |U_{\beta 2}|^2+|U_{\alpha 3}|^2 |U_{\beta 3}|^2\nonumber \\
&=&P(\bar{\nu}_{\beta} \to \bar{\nu}_{\alpha})+\bar{f}_{\rm reg}(|U_{\beta 2}|^2-|U_{\beta 1}|^2)
\end{eqnarray}
for antineutrinos.
Hence Eqs.~(\ref{eVO}) and (\ref{ebarVO}) become~\footnote{Here we do not discuss Earth matter effects to $F_x$. }
\begin{eqnarray}
F_e             & = &  (0.55-\rho) F_e^0 + (0.45+\rho) F_x^0, \label{eVO_mod}  \\
F_{\bar{e}}  & = &  (0.55-\bar{\rho}) F_{\bar{e}}^0 + (0.45+\bar{\rho})F_x^0,    \label{ebarVO_mod}  
\end{eqnarray}
with $\rho=f_{\rm reg}(|U_{e1}|^2-|U_{e2}|^2)$
and  $\bar{\rho}=\bar{f}_{\rm reg}(|U_{e1}|^2-|U_{e2}|^2)$.
For MSW-NO with Earth matter effects, the flux spectra of $\nu_e$ and $\bar{\nu}_e$ are given by  
\begin{eqnarray} 
F_e            & = &  F^0_x,   \label{eNH_m}  \\
F_{\bar{e}} & = &  (\cos^2\theta_{12}-\bar{f}_{\rm reg}) F^0_{\bar{e}} + (\sin^2\theta_{12}+\bar{f}_{\rm reg}) F^0_{\bar{x}}, \label{ebarNH_m}  
\end{eqnarray}
while their flux spectra in MSW-IO are
\begin{eqnarray} 
F_e            & = & (\sin^2\theta_{12}+f_{\rm reg}) F^0_e + (\cos^2\theta_{12}-f_{\rm reg}) F^0_x,  \label{eIH_m} \\
F_{\bar{e}} & = & F^0_{\bar{x}}. \label{ebarIH_m}  
\end{eqnarray}
Finally, FE scenario implies that all three neutrino mass eigenstates arrive on Earth in equal numbers. Since $F_\alpha (E,t)=F_i(E,t)P(\nu_i\to \nu_\alpha)$ with
$F_i(E,t)$ the flux spectrum of $i$th neutrino mass eigenstate arriving on the Earth, one still has $F_e=F_{\mu}=F_{\tau}$ as a consequence of $F_1=F_2=F_3$.
This follows from the condition $\sum_i P(\nu_i\to \nu_\alpha)=1$ despite the Earth matter effects given by Eq.~(\ref{nu_regeneration}) have been included in transition probabilities. 
Similarly we have $F_{\bar{e}}=F_{\bar{\mu}}=F_{\bar{\tau}}$.         

Numerically, for a neutrino path length $L=8000$ km inside the Earth, $f_{\rm reg}$ oscillates between $0$ and $0.06$~\cite{Liao:2016uis} for most probable energies of $\nu_e{\rm Ar}$ events during the accretion phase, $\sim 15$ MeV, as given by~\cite{Capozzi:2018rzl} based upon simulation G. Since the most probable energies for $\nu_e{\rm Ar}$ events for $t\leq 0.1$s considered here are comparable, one expects the maximum amplitude of $f_{\rm reg}$ is also about $0.06$. For the extreme case with $L=12000$ km, $f_{\rm reg}$ does not increase much.  Hence, to make simple estimations for Earth matter effects, we take an average value for $f_{\rm reg}$, i.e., $f_{\rm reg}=0.03$. 
Since $\rho=f_{\rm reg} (|U_{e1}|^2-|U_{e2}|^2)=f_{\rm reg}\cos 2\theta_{12}+{\mathcal O}(\sin^2\theta_{13})$, we have $\rho\simeq 0.01$ by using the best-fit value of $\sin^2\theta_{12}$ mentioned before. To estimate $\bar{f}_{\rm reg}$, 
we note that simulation G also predicts the most probable energies of IBD events to be $\sim 15$ MeV for the accretion phase~\cite{Capozzi:2018rzl}.   
From the result of~\cite{Liao:2016uis} with  $L=8000$ km, $\bar{f}_{\rm reg}$ oscillates between $-0.06$ and $0$ for $E_{\nu}$ around $15$ MeV. Following the previous argument for $f_{\rm reg}$, it is reasonable to take $\bar{f}_{\rm reg}=-0.03$ for making estimations. This leads to $\bar{\rho}\simeq -0.01$.

In Table~\ref{tab:Garching_matter}, we summarize the values of ${\mathcal A}^{i,{\rm Ar}}$, ${\mathcal A}^{i,{\rm IBD}}$, and ${\mathcal R}^i$ with and without Earth matter effects where the SN neutrino emission is based upon simulation G, the SN distance is taken as $5$ kpc, and
the value for ${\mathcal A}^{i,{\rm IBD}}$ is that expected in JUNO detector. 
We note that the Earth matter effects to  ${\mathcal A}^{\rm FE,Ar}$, ${\mathcal A}^{\rm FE,IBD}$, and ${\mathcal R}^{\rm FE}$ vanish as we have argued earlier. It is seen that ${\mathcal A}^{{\rm IO, Ar}}$ and ${\mathcal R}^{\rm IO}$ are most affected by Earth matter effects. However, these effects are still negligible for our analysis.   

\begin{table}[H]%
	\centering%
	\begin{tabular}{ccc}
		\toprule
		\hspace{4em} & w/o Earth matter effect \hspace{2em} & w/ Earth matter effect \\
		\cline{2-3}
		& \multicolumn{2}{c}{$\mathcal{A}^{i,\text{Ar}}$} \\
		\cline{2-3}
		VFT & $0.70 \pm 0.01$ & $0.70 \pm 0.01$\\
		NO & $0.52 \pm 0.01$ & $0.52 \pm 0.01$\\
		IO & $0.61 \pm 0.01$ & $0.63 \pm 0.01$\\
		\cline{1-3}
		& \multicolumn{2}{c}{$\mathcal{A}^{i,\text{IBD}}$} \\
		\cline{2-3}
		VFT & $0.49 \pm 0.01$ & $0.49 \pm 0.01$\\
		NO & $0.48 \pm 0.01$ & $0.48 \pm 0.01$\\
		IO & $0.52 \pm 0.01$ & $0.52 \pm 0.01$\\
		\cline{1-3}
		& \multicolumn{2}{c}{${\mathcal R}^i$} \\
		\cline{2-3}
		VFT & $1.44 \pm 0.01$ & $1.43 \pm 0.01$\\
		NO & $1.08 \pm 0.01$ & $1.08 \pm 0.01$\\
		IO & $1.19 \pm 0.01$ & $1.21 \pm 0.01$\\
		\botrule
	\end{tabular}
	\caption{Values of ${\mathcal A}^{i,{\rm Ar}}$, ${\mathcal A}^{i,{\rm IBD}}$, and ${\mathcal R}^i$ with and without corrections from Earth matter effects with Garching simulation of SN neutrino emissions. Here $d=5 \ {\rm kpc}$ is taken and the value for  ${\mathcal A}^{i,{\rm IBD}}$ 
	is that expected in the JUNO detector.}
	\label{tab:Garching_matter}%
\end{table}%

\section{Summary and Conclusions}

We have proposed to use the time evolution of SN neutrino event rates during the neutronization burst to test MSW effects occurring in SN neutrino propagation. 
The non-MSW scenarios for comparisons are the incoherent flavor transition probability for neutrino propagation in the vacuum and the flavor equalization
induced by fast flavor conversions. 
The event rates for various flavor transition scenarios are calculated with SN neutrino emissions extracted from simulations of four groups~\cite{Huedepohl:2009wh,Burrows:2019rtd,Nakazato:2012qf,Fischer:2015sll}. The behaviors of neutrino emissions in these four simulations are analyzed. 

To characterize the neutronization peak of $\nu_e$ flux in MSW-IO, FE, and VFT scenarios, we define cumulative time distribution $K^{i,\rm{Ar}}(t^*)$ of SN $\nu_e$ event in a liquid argon detector for $t^*$ between $0$ and $0.1$s as in Eq.~(\ref{cum_Ar}). 
To further quantify the ordering of $K^{i,\rm{Ar}}(t^*)$ for different flavor transition scenarios, we define the integral $\mathcal{A}^{i,\rm{Ar}}$ as given by Eq.~(\ref{Icd_Ar}). 
It is seen from left panels of Fig.~\ref{inte_cumu} that, for $d=1 \ {\rm kpc}$ and $5 \ {\rm kpc}$, $\mathcal{A}^{\rm{NO,Ar}}$ is distinguishable from all of $\mathcal{A}^{\rm{VFT,Ar}}$,  $\mathcal{A}^{\rm{FE,Ar}}$ and, $\mathcal{A}^{\rm{IO,Ar}}$ in the DUNE detector given statistical uncertainties and simulation dependencies. However neither $\mathcal{A}^{\rm{VFT,Ar}}$ and $\mathcal{A}^{\rm{IO,Ar}}$  nor $\mathcal{A}^{\rm{FE,Ar}}$ and $\mathcal{A}^{\rm{IO,Ar}}$ can be separated. To discriminate MSW-IO from VFT or FE, we invoke IBD events caused by SN $\bar{\nu}_e$ flux and measured by JUNO or HyperK detectors, i.e., we define the integral $\mathcal{A}^{i,\rm{IBD}}$ as in the case of liquid argon detector.  
 
We observed that $\mathcal{A}^{\rm{IO,IBD}} > \mathcal{A}^{\rm{VFT,IBD}}$ while $\mathcal{A}^{\rm{IO,Ar}} < \mathcal{A}^{\rm{VFT,Ar}}$. Additionally we also observed that $\mathcal{A}^{\rm{IO,IBD}} > \mathcal{A}^{\rm{FE,IBD}}$ 
while $\mathcal{A}^{\rm{IO,Ar}} < \mathcal{A}^{\rm{FE,Ar}}$. We have therefore taken advantages
of such orderings and defined the ratio $\mathcal{R}^i$ in Eq.~(\ref{ratio_Icd}). It is clearly seen from Fig.~\ref{rIcd} that, for all three chosen SN distances, MSW-IO can be clearly separated from VFT by combining either DUNE and JUNO measurements or DUNE and HyperK measurements.
In the same way, MSW-IO is separable from FE for $d=1 \ {\rm kpc}$ due to small statistical uncertainties. Finally it is interesting to see that for all three SN distances, VFT and FE are clearly separable. 
 We also argued that Earth matter effects are negligible in our analysis. Hence our method is effective for determining whether MSW effects indeed occur or not in the propagation of SN neutrinos. 

In conclusion, the combined observations of $\nu_e{\rm Ar}$ and IBD events are imperative for discriminating MSW from VFT or FE flavor transitions in the propagation of SN neutrinos
during the neutronization burst era. We have seen that DUNE detector can in general separate
MSW-NO from MSW-IO, VFT and FE while the latter three can be further discriminated with the IBD events of JUNO or HyperK detectors included for a combined analysis.

\section*{Acknowledgements}
We thank M.-R. Wu for useful discussions.   
The work is supported by National Science and technology Council, Taiwan under Grants No.~107-2119-M-009-017-MY3 and No.~110-2112-M-A49-006.

\newpage
\appendix
\section{Statistical Error Propagation in Cumulative Time Distribution $K(t^*)$ and Its Time Integration ${\mathcal A}$}
As we have already discussed in our previous context, the cumulative time distribution $K(t^*)$ is defined as the ratio between two event numbers, the number of events $N_{t^*}$ within a time interval $0\leq t^{\prime} \leq t^*$, and the total event number $N_T$ within the time interval $0\le t^{\prime}\le T$, where $T=100$ ms in our study:
\begin{align}
	K(t^*) &\equiv \frac{N_{t^*}}{N_T} ,\\
	N_{t^*}&=\int_0^{t^*}\frac{dN}{dt^{\prime}}dt^{\prime} ,\nonumber\\
	N_{T}&=\int_0^{T} \frac{dN}{dt^{\prime}}dt^{\prime} .\nonumber
	\label{K_def}
\end{align}
For simplicity in notations, we shall replace $t^*$ with $t$ hereafter. 

By definition, the variances for $N_t$ and $N_T$ are given by
\begin{align}
	\sigma_{t}^2 &\equiv \lim_{L\rightarrow\infty} \left[ \frac{1}{L} \sum_{i=0}^{L} (N_{t,i}-N_t)^2 \right] ,\\
	\sigma_{T}^2 &\equiv \lim_{L\rightarrow\infty} \left[ \frac{1}{L} \sum_{i=0}^{L} (N_{T,i}-N_T)^2 \right].
\end{align}
Where $L$ is the number of repeated measurements, $N_t$ and $N_{t,i}$ stand for the mean of measured event numbers and the event number of  
%$i\in [0\mathrel{{.}\,{.}}\nobreak N]$ 
$i$th measurement, respectively. Since $K(t)$ is a function of two variables $N_t$ and $N_T$, the standard deviation of $K(t)$ can be derived from the well-known error propagation formula, i.e., 
\begin{align}
	\sigma_{K(t)} = K(t)\times\sqrt{ \frac{\sigma_t^2}{N_t^2} + \frac{\sigma_{T}^2}{N_{T}^2} - 2\frac{\sigma_{tT}^2}{N_tN_T}}, \label{eq:CDFsd} 
\end{align}
where $\sigma_{t,T}^2$ is the covariance between the variables $N_t$ and $N_T$, which is given by
\begin{align}
	\sigma_{tT}^2 \equiv \lim_{L\rightarrow\infty} \left[ \frac{1}{L} \sum_{i=0}^{L} (N_{t,i}-N_t)(N_{T,i}-N_T) \right]. \label{eq:covariance}
\end{align}
 In addition, one can see that $N_t$ is actually involved in $N_T$ since
\begin{align}
	N_{T}&=\int_0^{T}\frac{dN}{dt^{\prime}}dt^{\prime} = \int_0^{t}\frac{dN}{dt^{\prime}}dt^{\prime} + \int_t^{T}\frac{dN}{dt^{\prime}}dt^{\prime} ,\nonumber\\
	&= N_{t} +  \int_t^{T}\frac{dN}{dt^{\prime}}dt^{\prime}, \nonumber \\
	&=N_{t}+N_{t,T},
\end{align} 
with
\begin{align}
N_{t,T}\equiv \int_t^{T}\frac{dN}{dt^{\prime}}dt^{\prime}.
\end{align} 
Therefore, the covariance term cannot be neglected. %Furthermore, a simple guess can be made on the covariance $\sigma_{tT}^2$ from the above relation. It is clear that the common contribution to $N_t$ and $N_T$ is $N_t$ itself, 
In fact one can show that $\sigma_{t,T}^2= \sigma_{t}^2$. 
To prove this, we
start from  Eq.~(\ref{eq:covariance}) and rearrange the right-hand side of the equation. Hence,
\begin{align}
	 \sigma_{t,T}^2 &\equiv \lim_{L\rightarrow\infty} \left[ \frac{1}{L} \sum_{i=0}^{L} (N_{t,i}-N_t)(N_{T,i}-N_T) \right] ,\nonumber\\
	&= \lim_{L\rightarrow\infty} \left[ \frac{1}{L} \sum_{i=0}^{L} (N_{t,i}-N_t)(N_{t,i} - N_t + N_{t,T,i} - N_{t,T}) \right] ,\nonumber\\
	&= \lim_{L\rightarrow\infty} \left[ \frac{1}{L}\sum_{i=0}^{L} (N_{t,i}-N_t)^2 \right] + \lim_{L\rightarrow\infty} \left[ \frac{1}{L}  \sum_{i=0}^{L}(N_{t,i}-N_t)(N_{t,T,i}-N_{t,T}) \right] ,\nonumber\\
	&= \sigma_{t}^2.
	\label{covariance_proof}
\end{align}
The last equality  holds because the first term is $\sigma_t^2$ by definition while the second term vanishes because $N_t$ and $N_{t,T}$ are measurements in different time windows, that do not correlate with each other.
%we should expect to find equal distributions of positive and negative values for the second term in a large random selection of observations, therefore, the second term vanishes and remains only the first term, which is $\sigma_{t}^2$ by definition.

Neutrino events are rare due to rather small interaction cross sections. Therefore, we should expect the statistical uncertainty of the measurement follows Poisson distribution, i.e., the standard deviation can be obtained directly from the mean event number $\sigma=\sqrt{N}$.
\begin{align}
	\sigma_{t,T}^2 &= \sigma_{t}^2 \overset{\mathcal{P}}{=} N_t ,\\
	\sigma_{t}^2 &\equiv \lim_{N\rightarrow\infty} \left[ \frac{1}{N}(N_{t,i}-N_t)^2 \right] \overset{\mathcal{P}}{=} N_t, \\
	\sigma_{T}^2 &\equiv \lim_{N\rightarrow\infty} \left[ \frac{1}{N}(N_{t,T}-N_T)^2 \right] \overset{\mathcal{P}}{=} N_T \label{eq.covariance_a}.
\end{align}
Here $\mathcal{P}$ stands for the Poisson statistics. By substituting the above equations for variance and covariance into Eq.~\eqref{eq:CDFsd}, we obtain
\begin{align}
	\sigma_{K(t)} &= K(t)\times\sqrt{ \frac{\sigma_t^2}{N_t^2} + \frac{\sigma_{T}^2}{N_{T}^2} - 2\frac{\sigma_{t,T}^2}{N_tN_T}} ,\nonumber\\
	&= K(t)\times \sqrt{ \frac{1}{N_t} - \frac{1}{N_T} }.
\end{align}
\newpage

To derive the variance of the time-integrated cumulative distribution $\mathcal{A}$, we recall the definition of $\mathcal{A}$,  
\begin{align}
	\mathcal{A} &\equiv \frac{1}{T}\int_{0}^{T} K(t) dt, 
\end{align}
where, as stated before, $t^*$ has been replaced by $t$ for simplicity. 
Since we have binned SN neutrino events with $5$ ms bin size for a total time period $T=100$ ms, the above integral is in fact a summation of $20$ terms, i.e., 
\begin{align}
	\frac{1}{T}\int_{0}^{T} K(t) dt &\equiv \frac{1}{T} \lim_{N\rightarrow\infty} \sum_{i=1}^{N} K(t_i) \Delta t \\
	&\rightarrow \frac{1}{T} \sum_{i=1}^{20} K(t_i) \Delta t ,\\
	&= \frac{5 \text{ ms}}{100 \text{ ms}} \sum_{i=1}^{20} K(t_i).
\end{align}
%The interval of each time bin and the total time has been chosen as $5$ ms and $T=100$ ms respectively in our analysis, therefore, the number of bin $N=20$, i.e., $i$ from $0$ to $19$, in the last two lines of above formula. 
From Eq.~(\ref{K_def}), we may write 
\begin{align}
K(t_i)=\sum_{l=1}^i f_l,
\end{align}
with
\begin{align}
f_l=\frac{1}{N_T}\int_{t_{l-1}}^{t_l}\frac{dN}{dt^{\prime}}dt^{\prime}\equiv \frac{N_l}{N_T}.
\end{align}
Therefore, 
\begin{align}
\sum_{i=1}^{20}K(t_i)=\sum_{m=1}^{20}(21-m)f_m,
\end{align}
and the variance of $\mathcal{A}$ is given by
\begin{align}
\sigma_{\mathcal A}^2=(\frac{5{\rm ms}}{100{\rm ms}})^2\left(\sum_{m=1}^{20}(21-m)^2\sigma_{f_m}^2+\sum_{1\leq l<m}2(21-l)(21-m){\rm cov}(f_l,f_m)\right).    
\end{align}
For Poisson distribution, it is easy to show that
\begin{align}
\sigma^2_{f_l}=\frac{f_l(1-f_l)}{N_T}, \ {\rm cov}(f_l,f_m)=-\frac{f_lf_m}{N_T}.    
\end{align}
With these results, $\sigma_{\mathcal A}$ can be readily calculated. 
\end{document}